\documentclass[twocolumn]{aastex62}

\usepackage[natbib]{}
\usepackage{graphicx}
\usepackage{bm}
\usepackage{mathtools}
\usepackage{siunitx}
\usepackage{enumitem}
\usepackage{multirow}

\usepackage[flushleft]{threeparttable}

\shorttitle{AMISS III: CO Spectral Line Energy Distributions}
\shortauthors{Keenan et al.}

\defcitealias{keenan+24a}{Paper~I}
\defcitealias{keenan+24b}{Paper~II}

\begin{document}

\title{The Arizona Molecular ISM Survey with the SMT:\\
The Diverse Carbon Monoxide Line Ratios and Spectral Line Energy Distributions of Star Forming Galaxies}

\correspondingauthor{R. P. Keenan}
\author[0000-0003-1859-9640]{Ryan P. Keenan}
\affiliation{Max-Planck-Institut für Astronomie, Königstuhl 17, D-69117 Heidelberg, Germany}
\affiliation{Steward Observatory, University of Arizona, 933 North Cherry Avenue, Tucson, AZ 85721, USA}
\email{keenan@mpia.de}

\author[0000-0002-2367-1080]{Daniel P. Marrone}
\affiliation{Steward Observatory, University of Arizona, 933 North Cherry Avenue, Tucson, AZ 85721, USA}

\author[0000-0002-3490-146X]{Garrett K. Keating}
\affiliation{Center for Astrophysics, Harvard \& Smithsonian, 60 Garden Street, Cambridge, MA 02138, USA}

\begin{abstract}

The carbon monoxide (CO) spectral line energy distributions (SLEDs) of galaxies contain a wealth of information about conditions in their cold interstellar gas. Here we use galaxy-scale observations of the three lowest energy CO lines to determine SLEDs and line ratios in a sample of 47 nearby, predominantly star forming galaxies. We find systematic trend of higher gas excitation with increasing star formation rate (SFR) and SFR surface density ($\Sigma_{\rm SFR}$), with the range of variations being even larger than predicted by simulations. Power law fits of the CO line ratios as a function of SFR and $\Sigma_{\rm SFR}$ provide a good description of the trends seen in our sample and also accurately predict values for a wide range of galaxy types compiled from the literature. Based on these fits, we provide prescriptions for estimating CO(1--0) luminosities and molecular gas masses using CO(3--2) or CO(2--1) in cases where CO(1--0) is not observed directly. We compare our observed SLEDs with molecular cloud models in order to examine how the physical properties of cold gas vary across the galaxy population. We find that gas conditions in star forming and starburst galaxies lie on a continuum with increasing gas density in more actively star forming systems.

\end{abstract}



\section{Introduction} \label{sec:intro}

Star formation is regulated by the conditions in the cold, predominantly molecular phase of the interstellar medium \citep[ISM;][]{kennicutt+12,schinnerer+24}. The available reservoir of molecular gas, combined with the efficiency at which this is converted into stars, sets a galaxy's star formation rate. These properties also influence the location of a galaxy relative to the tight correlation between stellar mass and SFR followed by most star forming galaxies 
\citep[known as the star forming galaxy main sequence;][]{saintonge+17,saintonge+22}. Galaxies with high star formation efficiency (SFE) tend to lie above the main sequence, with the most extreme starburst galaxies exhibiting SFEs roughly ten times higher than main sequence galaxies \citep{sargent+14}. Measuring physical conditions -- the density and temperature distributions -- of the ISM is necessary to refine our understanding of the star formation process, both in the nearby universe and across cosmic time \citep{daddi+15,liu+21,valentino+20,boogaard+20,weiss+05}.

Lower rotational levels of CO are easily populated in the density and temperature conditions prevailing in molecular clouds, producing a ladder of bright spectral lines starting at 115.27 GHz (the $J=1\rightarrow0$ transition, CO(1--0)) and extending into the far-infrared. The ratios of flux arising from the different transitions depends on the density, temperature, and optical depth of the clouds, with higher-$J$ transitions generally becoming brighter in hotter and denser gas. The distribution of energy across these CO transitions can be used to determine physical conditions in the molecular ISM \citep{leroy+17}.

The spectral line energy distributions (SLEDs) of IR-selected starburst galaxies show bright emission throughout the low-$J$ CO transitions \citep{kamenetzky+14,papadopoulos+12,montoya-arroyave+23}, leading to a small dynamic range in the CO line ratios. On the other hand, disky galaxies near the main sequence show a wider range of CO line excitation \citep{keenan+24b,denbrok+21,lamperti+20}. Significant insight about the ISM of these galaxies may therefore be gained from just the low-$J$ CO transitions \citep{denbrok+23b,leroy+22,gong+20,penaloza+17,penaloza+18}.

Here, we use observations of the CO(1--0), CO(2--1), and CO(3--2) spectral lines to determine low-$J$ CO SLEDs of 47 $z\sim0$ galaxies. By identifying trends between the CO line ratios and galaxy properties such as SFR and star formation rate surface density ($\Sigma_{\rm SFR}$), we link observable indicators of molecular ISM conditions to the global evolutionary state of the host galaxy. We then compare the observed CO line ratios to models of molecular clouds with realistic density distributions in order to assess how the density and temperature of the molecular gas change as star formation in a galaxy becomes more intense.

Knowledge of the low-$J$ CO SLED is also valuable for determining molecular gas masses from observations of CO(2--1) or CO(3--2) when CO(1--0) -- the standard gas mass tracer \citep{bolatto+13} -- is not available. We therefore provide parameterizations of the the CO(3--2)/CO(1--0) and CO(2--1)/CO(1--0) luminosity ratios as functions of easily accessible galaxy properties.

This paper is part of a series describing the results of the Arizona Molecular ISM Survey with the SMT (AMISS), which was designed as a detailed study of the CO(1--0), CO(2--1) and CO(3--2) lines in a diverse sample of $z\sim0$ galaxies. Prior papers have described the survey design and execution \citep[][hereafter \citetalias{keenan+24a}]{keenan+24a}, explored the calibration of CO(2--1) as a molecular gas mass tracer \citep[][hereafter \citetalias{keenan+24b}]{keenan+24b}, and examined how the choice of molecular gas tracer affects the measured slope of the Kennicutt-Schmidt relation (Samboco \& Keenan, subm.).

In the remainder of this paper we introduce our sample and data (Section~\ref{sec:survey}), characterize how the low-$J$ CO SLEDs (Section~\ref{sec:sled}) and line ratios (Section~\ref{sec:ratios}) depend on other galaxy properties, and then use these results to evaluate how the physical conditions of the ISM change in conjunction with increasing star formation (Section~\ref{sec:ism_conditions}). Throughout this paper we assume a a flat $\Lambda$CDM cosmology with $H_0=70$ and $\Omega_m=0.3$. We use $\log$ to denote base 10 logarithms, and we quote CO line ratios as the ratios of Rayleigh-Jeans brightness temperatures. SFRs are derived assuming a \citet{chabrier03} stellar initial mass function.


\section{Sample and Data}\label{sec:survey}

\begin{figure}
    \centering
    \includegraphics[width=.45\textwidth]{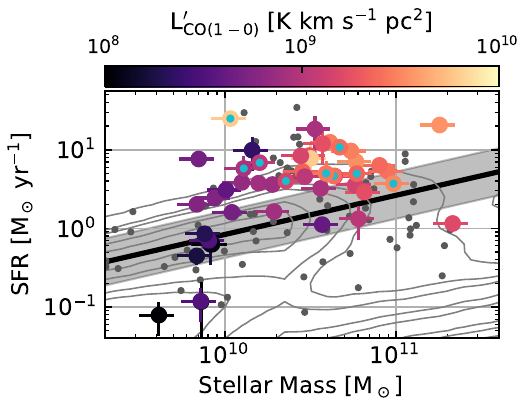}
    \caption{The distribution of our sample in stellar mass, SFR, and CO(1--0) luminosity (color-axis) is shown by large points. Galaxies with CO(3--2) data from \citet{lamperti+20} are indicated by blue fill in the markers. Additional AMISS galaxies included in \citetalias{keenan+24b}'s analysis of the $r_{21}$ ratio are shown by smaller, gray points. The black line and gray filled region show the star forming main sequence of \citet{speagle+14}, and gray contours show the distribution of SFR at a given mass for all $z<0.05$ SDSS galaxies. }
    \label{fig:sample}
\end{figure}

\begin{figure*}
    \centering
    \includegraphics[width=\textwidth]{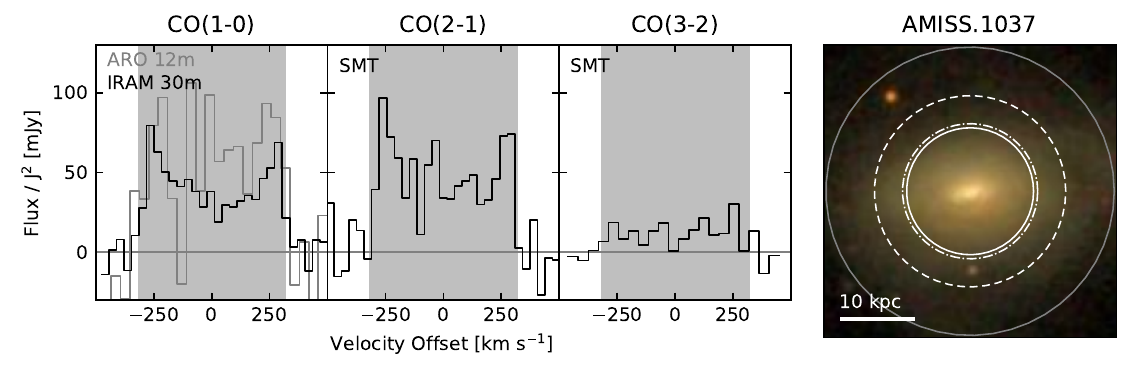}
    \caption{CO spectra and an SDSS composite image for AMISS.1037. For CO(1--0) we show spectra from both the IRAM 30m (black) and ARO 12m (gray). In the right panel we superimpose the half-power beam sizes for each telescope on the optical image. Based on the optical profile we estimate that the IRAM CO(1--0) beam (solid white line) and SMT CO(3--2) beam (dash-dotted line) miss 22\% and 21\% of the total flux. The SMT beam for CO(2--1) (dashed line) misses 12\% of the flux. The IRAM~30m CO(1--0) flux is 21\% lower than ARO~12m CO(1--0) flux measured in a larger aperture (gray circle), indicating that these estimates are reasonable. We correct CO luminosities and line ratios for this missing flux; matched CO(1--0) and CO(3--2) apertures mean that these corrections have little effect on the $r_{31}$ ratio.}
    \label{fig:example_galaxy}
\end{figure*}

We base our analysis on the public, multi-$J$ CO line catalogs from AMISS. Details of the survey design, execution and data reduction can be found in \citetalias{keenan+24a}. In brief, the data consist of galaxy-scale, single-pointing spectra combining archival CO(1--0) data from the IRAM 30m Telescope \citep[xCOLD~GASS;][]{saintonge+17} with new CO(2--1) and CO(3--2) observations from the Arizona Radio Observatory's Submillimeter Telescope (SMT). The final survey included CO(2--1) observations of 176 galaxies (130 detections with signal to noise ratio above three)  and CO(3--2) observations of a subset of 45 (34 detections). CO(2--1) targets were selected to provide a representative sample of the $M_*>10^9$~M$_\odot$ galaxy population at $z\sim0$; CO(3--2) targets consisted of the subset of these bright enough to be detected with the SMT's less sensitive 0.8~mm receiver. The complete AMISS dataset, including spectral line catalog and 277 observed spectra, is available via Zenodo (\href{https://doi.org/10.5281/zenodo.11085052}{doi:10.5281/zenodo.11085052}).

Here we select all 42 galaxies for which AMISS has reliable CO(1--0) through CO(3--2) measurements and ancillary data. Additionally, eighteen AMISS targets were independently observed in CO(3--2) by \citet{lamperti+20} using the James Clerk Maxwell Telescope (JCMT). We supplement our CO(3--2) data with these results, choosing the measurement with the lower statistical noise when both JCMT and SMT CO(3--2) data are available, bringing our final sample to 47 galaxies. In \citetalias{keenan+24a} we verified that CO(3--2) aperture corrected luminosities derived from the AMISS and \citet{lamperti+20} datasets are consistent among 10 objects with CO(3--2) data from both projects. All of our targets are detected ($\geq 3\sigma$) in CO(1--0) and CO(2--1), while for CO(3--2) we obtained 36 detections and 11 upper limits. 

Figure~\ref{fig:sample} shows the distribution of the sample in stellar mass, SFR and CO(1--0) luminosity. The galaxies tend to lie on and above the star forming main sequence, a consequence of selecting CO(3--2) targets from those galaxies with bright CO(1--0) or CO(2--1) emission. 

We correct the CO measurements to account for flux falling outside the primary beam of the single-pointing observations. As targets were selected to be comparable to or smaller than the beam, these corrections are generally small. Our estimates of the median missing flux are 19\% for CO(1--0), 10\% for CO(2--1), and 21\% for CO(3--2). Among our sample, AMISS.1037 has the median CO(3--2) aperture correction; as an example, we show the CO spectra and an SDSS composite image of this galaxy superimposed with the beams used to measure each line in Figure~\ref{fig:example_galaxy}. We are able to directly estimate the missing CO(1--0) flux for this galaxy based on a second CO(1--0) measurement in a larger aperture (see below), and find it to be 21\%, in near-perfect agreement with our model-based aperture correction of 20\%.

We derive line luminosities as
\begin{equation}
    L_{\rm CO}^\prime = 3.25\times10^7 S\Delta v \frac{D_L^2}{(1+z)\nu^2_{\rm CO}} ,
\end{equation}
where $L_{\rm CO}^\prime$ has units of K~km~s$^{-1}$~pc$^2$, $S\Delta v$ is the aperture corrected, integrated flux in Jy~km~s$^{-1}$, $D_L$ is the luminosity distance in Mpc, $z$ is redshift, and $\nu_{\rm CO}$ is the rest frequency of the CO line in GHz. We compute CO line ratios as 
\begin{equation}
    r_{jk} = \frac{L_{{\rm CO}(j-(j-1))}^\prime}{L_{{\rm CO}(k-(k-1))}^\prime} .
\end{equation}
The median statistical uncertainties are 15\%, 23\%, and 29\% for $r_{21}$, $r_{31}$, and $r_{32}$ respectively. Additional systematic uncertainty is introduced by flux calibration and aperture corrections.
We estimate the median systematic uncertainties to be 12\%, 11\%, and 8\% for $r_{21}$, $r_{31}$, and $r_{32}$ (see \citetalias{keenan+24a} for details). 

We derive molecular gas masses from CO(1--0) luminosities as
\begin{equation}
    M_{\rm mol} = \alpha_{\rm CO} L^\prime_{\rm CO(1-0)}\,
\end{equation}
where $\alpha_{\rm CO}$ is the CO to molecular mass conversion factor \citep{bolatto+13}. While many of our galaxies lie above the main sequence, they are in almost all cases disk galaxies with IR luminosities $\lesssim10^{11}\,{\rm L_\odot}$, and therefore we adopt a Milky Way-like conversion factor of $\alpha_{\rm CO}=4.3$~M$_\odot$~(K~km~s$^{-1}$~pc$^2$)$^{-1}$ (which includes a 36\% metal fraction). Adopting a ``ULIRG'' conversion factor for the two IR-luminous merging galaxies in our sample would not significantly alter our results. For 37 of our targets (79\%), we have a second CO(1--0) measurement from the Arizona Radio Observatory's 12m ALMA Prototype Antenna. For these galaxies, we derive CO line ratios and SLEDs using the IRAM 30m CO(1--0) data, but derive galaxy-integrated CO(1--0) luminosities and molecular gas masses using the 12m CO(1--0) data. This allows us to study correlations between the CO line ratios and molecular gas masses using statistically independent measurements.

We draw ancillary information -- including SFRs, stellar masses, and optical sizes -- from the public xCOLD~GASS catalog \citep{saintonge+17}. The catalog provides uncertainties for SFRs from \citet{janowiecki+17}, but does not include errors for most other quantities. Where necessary, we assume typical errors of 0.1~dex on stellar masses \citep{salim+16} and 0.2" for optical radii. We compute SFR and molecular gas mass surface densities as half of the global SFR/gas mass divided by the surface area within the optical half light radius.


\begin{figure*}
    \centering
    \includegraphics[width=\textwidth,trim={0 9mm 0 0},clip]{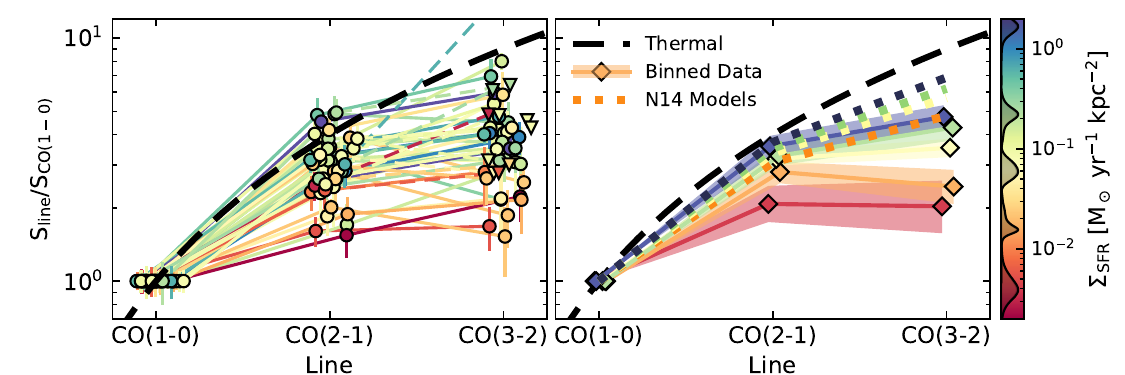}
    \includegraphics[width=\textwidth,trim={0 0 0 2.5mm},clip]{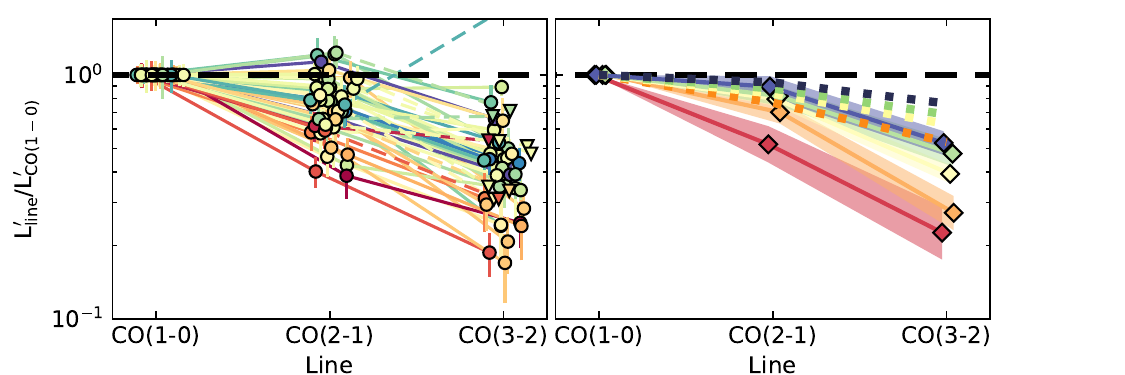}
    \caption{SLEDs for our galaxy sample normalized to CO(1--0). We plot the data in terms of both flux (top row) and line luminosity (bottom row). Left: SLEDs of individual galaxies. Points with error bars show detected CO lines and $1\sigma$ uncertainties, while downward triangles (connected to dashed lines) indicate $2\sigma$ upper limits for undetected CO(3--2). Galaxies are color coded according to $\Sigma_{\rm SFR}$, a proxy for the intensity of the interstellar radiation field in the galaxy. The black dashed line shows the expected scaling for optically thick emission in the Rayleigh-Jeans regime under LTE conditions.
    Right: SLEDs in five bins of $\Sigma_{\rm SFR}$. Solid lines and diamonds show medians for the binned galaxies, dotted lines show the predictions of \citet{narayanan+14} (for the four highest $\Sigma_{\rm SFR}$ bins only, as their model is not defined for the lowest $\Sigma_{\rm SFR}$ bin). Shaded regions around the median SLEDs show the 1$\sigma$ uncertainty determined by repeatedly perturbing each luminosity and $\Sigma_{\rm SFR}$ according to its uncertainty and recomputing the bins. The black curves enclosing grayed filled regions on the color bar show the distribution of the median $\Sigma_{\rm SFR}$ of the bins in the re-draws.}
    \label{fig:sled}
\end{figure*}

\section{CO Spectral Line Energy Distributions}\label{sec:sled}

Figure~\ref{fig:sled} presents the low-$J$ CO SLEDs for our 47 galaxies, normalized to CO(1--0) and color coded according to SFR surface density ($\Sigma_{\rm SFR}$) -- a proxy for the intensity of the interstellar radiation field \citep{narayanan+14,daddi+15}. There is significant scatter in our measurements, driven by uncertainties of the line ratios for individual galaxies. However, there is a general trend that galaxies with stronger radiation fields have higher CO line ratios and emission closer to that expected for thermalized excitation in an optically thick gas (black dashed line).

To show the typical trend more clearly we group galaxies into five bins evenly spaced in $\log \Sigma_{\rm SFR}$ and compute the median SLED of each bin. The results are shown in the right panel of Figure~\ref{fig:sled}, and compared to the prescription presented by \citet{narayanan+14} based on hydrodynamical simulations of disk and merging galaxies. The observed SLEDs show significantly greater variation as a function of $\Sigma_{\rm SFR}$ than is predicted by the simulations. At a given $\Sigma_{\rm SFR}$, the CO excitation in our sample is lower than predicted by the model, especially at the $J=3$ level. At low $\Sigma_{\rm SFR}$ our galaxies differ from the model SLED by a factor of two. 

These results indicate that the low-$J$ CO transitions are quite sensitive to changes in the state of the ISM in typical star forming galaxies. This is consistent with conclusions of recent studies of nearby star forming galaxies \citep{leroy+22,lamperti+20,denbrok+21,denbrok+23b,denbrok+25}, but differs from earlier findings for IR-selected samples, where variations in the SLED were primarily apparent in the $J_u\ge4$ transitions \citep{greve+14,kamenetzky+16,liu+15}.

\section{Low-$J$ CO Line Ratios}\label{sec:ratios}

\begin{figure*}
    \centering
    \includegraphics[width=\textwidth]{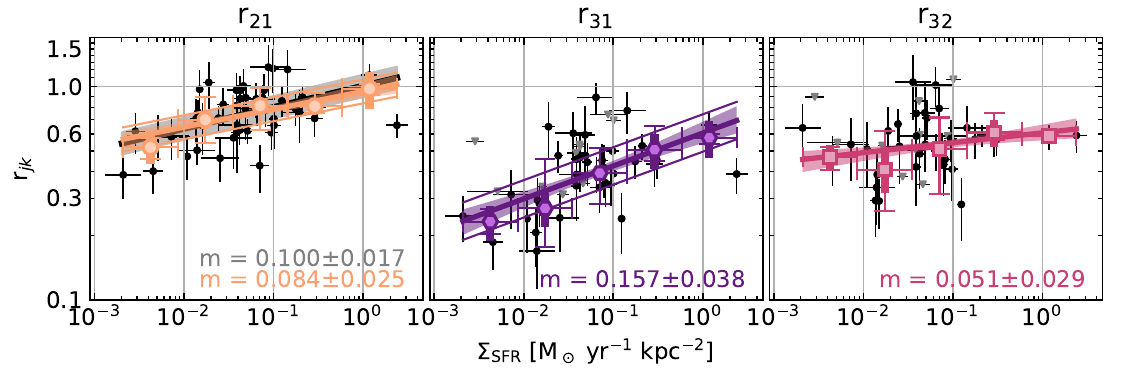}
    \caption{Trends between the CO line ratios and $\Sigma_{\rm SFR}$. Black points show measurements and $1\sigma$ uncertainties for individual objects; gray downward triangles show $2\sigma$ upper limits where CO(3--2) is not detected. Colored points show median values in equally spaced bins along the $x$-axis, with thin error bars showing the bin limits ($x$) and 16th-84th percentile range ($y$) and thick error bars showing the uncertainty of the median value in each bin. Thick lines and filled regions show the best fitting power law and its $1\sigma$ uncertainty, while thin lines show the best-fitting intrinsic scatter in the relation (the $2\sigma$ upper limit of 0.13~dex for the $r_{32}$ scatter is not shown). The black line and gray region in the left panel show the $r_{21}$--$\Sigma_{\rm SFR}$ fit from \citetalias{keenan+24b}.}
    \label{fig:trends}
\end{figure*}

\begin{figure*}
    \centering
    \includegraphics[width=\textwidth]{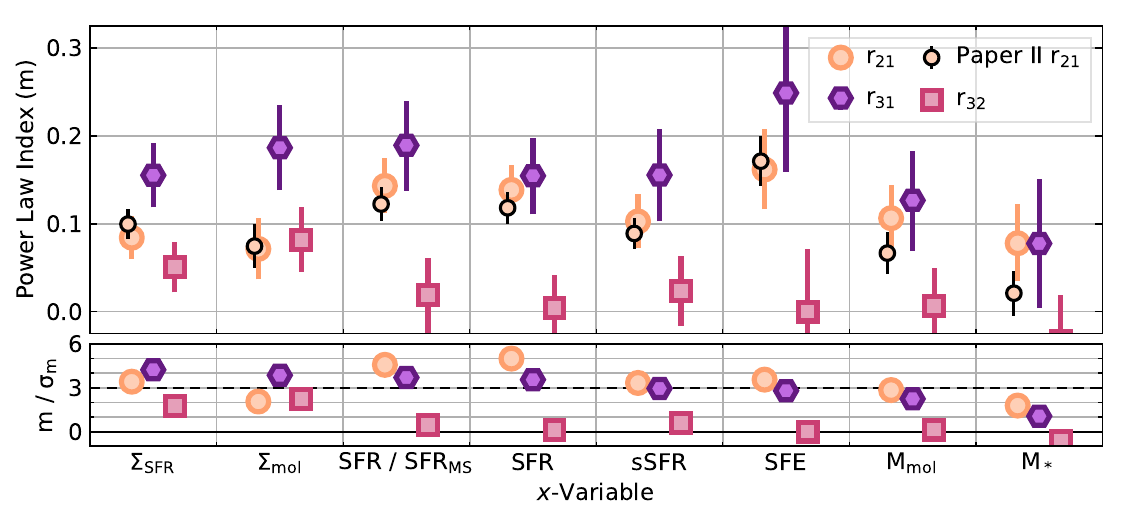}
    \caption{Top: Power law slopes ($m_{jk}$) between the $r_{21}$ (orange), $r_{31}$ (purple), or $r_{32}$ (pink) and a range of galaxy properties. Error bars show $1\sigma$ uncertainties. We also show $m_{21}$ found using the full AMISS sample in \citetalias{keenan+24b} (smaller markers with black outlines).
    Bottom: The significance by which the slopes differ from zero.
    The $x$-variables are ordered according to the significance of $m_{31}$.}
    \label{fig:trends_summary}
\end{figure*}

CO line ratios trace the physical conditions of molecular gas, and are correlated with global galaxy properties. We found in \citetalias{keenan+24b} that $r_{21}$ is significantly correlated with a number of quantities related to star formation -- SFR, $\Sigma_{\rm SFR}$, offset from the main sequence (${\rm SFR} / {\rm SFR}_{\rm MS}$), specific star formation rate (${\rm sSFR=SFR}/M_*$), and star formation efficiency (${\rm SFE}={\rm SFR}/M_{\rm mol}$). \citet{leroy+22} and \citet{lamperti+20} found evidence of correlations between $r_{31}$ and/or $r_{32}$ and subsets of these variables. \citetalias{keenan+24b} also showed weaker evidence of correlations between $r_{21}$ and molecular gas mass ($M_{\rm mol}$) and surface density ($\Sigma_{\rm mol}$). \citet{leroy+22} proposed that line ratios correlate with galaxy mass. In the following we consider how the three lowest-$J$ CO line ratios correlate with all of these galaxy properties.

We describe the relation between the CO line ratios and other galaxy properties using a power law of the form
\begin{equation}
	\log r_{jk} = m \log{x} + b + \epsilon (s)
\end{equation}
where $m$ and $b$ are the slope (or index) and normalization of the power law, and $\epsilon$ represents a log-normal intrinsic scatter of width $s$ \citep{leroy+22,keenan+24b}. We determine the best fitting parameters for each $r_{jk}$-$x$ pair using the Markov chain Monte Carlo fitting procedure described in \citet{hogg+10}, allowing for uncertainty in both $x$ and $y$ and accounting for covariant uncertainties where necessary. 

Full fit results are listed in Appendix~\ref{ap:fit_results}. To provide an illustration of these results, we show the line ratios and fits as a function of $\Sigma_{\rm SFR}$ in Figure~\ref{fig:trends}. We summarize the power law indices for all line ratio--galaxy property pairs in Figure~\ref{fig:trends_summary}. 

We quantify the likelihood of a correlation between $x$ and $r_{jk}$ in terms of the significance by which the fitted slope differs from zero.\footnote{We prefer this metric to correlation coefficients, since our data are noisy, do not follow a normal distribution in the $x$-variables, and have partially correlated $x$- and $y$-errors in some cases.} We find evidence ($m/\sigma_m\gtrsim3$) of non-zero slopes between $\log r_{21}$ and $\log r_{31}$ and all quantities except stellar mass (and only marginally for $M_{\rm mol}$). On the other hand, for $r_{32}$ only $\Sigma_{\rm SFR}$ and $\Sigma_{\rm mol}$ show modest evidence for positive correlations ($m/\sigma_m\gtrsim2$), while the remaining variables have power law slopes consistent with zero.

Theory and simulations suggest quantities such as $\Sigma_{\rm SFR}$, $\Sigma_{\rm mol}$ -- which normalize by galaxy size -- should be better predictors of CO line excitation than total SFR or $M_{\rm mol}$ \citep{narayanan+14,bournaud+15,penaloza+18,gong+20}. Observations of high-$J$ CO lines are often consistent with this expectation \citep{daddi+15,lamperti+20}. However in \citetalias{keenan+24b} we found no evidence that this normalization is necessary when considering only $r_{21}$. We can re-evaluate this conclusion in light of the multi-$J$ information presented here. In Figure~\ref{fig:trends_summary} we see that while $r_{21}$ and $r_{31}$ correlate with both SFR and $\Sigma_{\rm SFR}$, the $r_{32}$ ratio is flat with increasing SFR. On the other hand, $r_{32}$ shows a moderate, positive correlation with $\Sigma_{\rm SFR}$. Similarly, $r_{32}$ shows evidence of a correlation with $\Sigma_{\rm mol}$ but not  $M_{\rm mol}$.

These results are consistent with a picture in which a galaxy's CO(1--0) luminosity can include a contribution from a diffuse and/or cold, non-star-forming gas phase which is faint in the $J_u\ge2$ CO lines. While this is a common interpretation of the $r_{31}$ ratio \citep[e.g.][]{narayanan+05,komugi+07,miura+14,moromkuma-matsui+17}, CO(2--1) is often assumed to trace the same gas seen CO(1--0). However, the similarity between the $r_{21}$ and $r_{31}$ trends, along with the relatively flat trends in $r_{32}$ across all properties considered, may indicate that CO(3--2) and CO(2--1) are tracing more or less the same gas, while CO(1--0) emission captures a more diffuse phase of the ISM missed by all higher $J$ lines. In this picture, $r_{21}$ and $r_{31}$ can be interpreted as indicators of the fraction of molecular gas which is actively forming stars. 

Area-normalized quantities such as $\Sigma_{\rm SFR}$ are likely better predictors of the overall shape of the CO SLED. The moderate correlations of $r_{32}$ with $\Sigma_{\rm SFR}$ and $\Sigma_{\rm mol}$ suggest that these ratios are sensitive to the radiation field and gas density in the star-forming gas. Similar conclusions have been reached for higher $J$ line ratios such as $r_{52}$ \citep{liu+21,valentino+20,boogaard+20}.

In the remainder of this section, we provide updated prescriptions for estimating $r_{31}$ and $r_{32}$ when they cannot be measured directly, then we compare the trends found in our data with results from the literature.

\subsection{Prescriptions for Estimating Low-$J$ CO Line Ratios}\label{ss:presc}

Our power law fits allow us to provide prescriptions for estimating CO line ratios. This is often useful for deriving molecular gas masses from observations of $J_u\ge2$ CO transitions, which typically involves converting to a CO(1--0) luminosity using $r_{j1}$ and then applying the CO(1--0) conversion factor:
\begin{equation}
	M_{\rm mol} = \frac{\alpha_{\rm CO}}{r_{j1}} L_{{\rm CO}(j-(j-1))}^\prime
\end{equation}
In \citetalias{keenan+24b} we demonstrated that the standard practice of assuming a constant $r_{j1}$ introduces bias when studying the molecular gas properties of diverse galaxy samples \citep[see also][]{yajima+21}. A prescription that accounts for the variations of $r_{j1}$ is therefore necessary to reliably recover $M_{\rm mol}$ from CO(3--2) or CO(2--1) observations.

Because SFR is readily available for most galaxies, in \citetalias{keenan+24b} we recommend parameterizing $r_{21}$ in terms of SFR: 
\begin{equation}\label{eq:r21prescription}
    \log r_{21} =     
    \begin{dcases}
        0.12 \log {\rm SFR} - 0.19 & -3 \lesssim \log {\rm SFR} < 1.58 \\
        0.0 & 1.58 < \log {\rm SFR}
    \end{dcases}\,,
\end{equation}
where SFR is given in units of M$_\odot$~yr$^{-1}$. The upper limit of $r_{21}\le1.0$ is applied since it is difficult to produce super-thermal CO excitation on galaxy scales, and galaxies with extreme star formation seem to approach a limit of $r_{21}\sim1$ \citep{papadopoulos+12,montoya-arroyave+23}. The corresponding prescription for $r_{31}$ can be obtained from the fit in Table~\ref{tab:fits}:
\begin{equation}\label{eq:r31prescription}
    \log r_{31} =     
    \begin{dcases}
        0.15 \log {\rm SFR} - 0.49 & -3 \lesssim \log {\rm SFR} < 3.2 \\
        0.0 & 3.2 < \log {\rm SFR}
    \end{dcases}\,,
\end{equation}
again applying a limit of $r_{31}\leq1.0$.

The results of the preceding section suggest that $\Sigma_{\rm SFR}$ captures the full shape of the SLED better than SFR. In terms of $\Sigma_{\rm SFR}$ the low$-J$ CO line ratios can be parameterized as 
\begin{equation}\label{eq:r21sigsfr}
    \log r_{21} =     
    \begin{dcases}
        0.10 \log \Sigma_{\rm SFR} - 0.00 & -1 \lesssim \log \Sigma_{\rm SFR} < -0.04 \\
        0.0 & 0.04 < \log \Sigma_{\rm SFR}
    \end{dcases}\,,
\end{equation}
\begin{equation}\label{eq:r31sigsfr}
    \log r_{31} =     
    \begin{dcases}
        0.16 \log \Sigma_{\rm SFR} - 0.22 & -1 \lesssim \log \Sigma_{\rm SFR} < 1.40 \\
        0.0 & 1.40 < \log \Sigma_{\rm SFR}
    \end{dcases}\,.
\end{equation}
where $\Sigma_{\rm SFR}$ has units of M$_\odot$~yr$^{-1}$~kpc$^{-2}$.\footnote{We have used the $r_{21}$-$\Sigma_{\rm SFR}$ fit from \citetalias{keenan+24b}, which is based on a larger sample.}

The intrinsic variations in $r_{21}$ and $r_{31}$ are larger than the uncertainty in the fit parameters over the full SFR and $\Sigma_{\rm SFR}$ range of our data. We therefore recommend adopting uncertainties of 0.1~dex for $r_{31}$ values derived using Equations~\ref{eq:r31prescription} and \ref{eq:r31sigsfr} and 0.05~dex for $r_{21}$ derived from Equations~\ref{eq:r21prescription} and \ref{eq:r21sigsfr}. As we show in the next section, these fits successfully match not just the AMISS data, but also line ratios from a diverse literature sample.

\subsection{Low-$J$ CO Line Ratios in the Literature}\label{ss:litcomp}

\begin{figure}
    \centering
    \includegraphics[width=.49\textwidth]{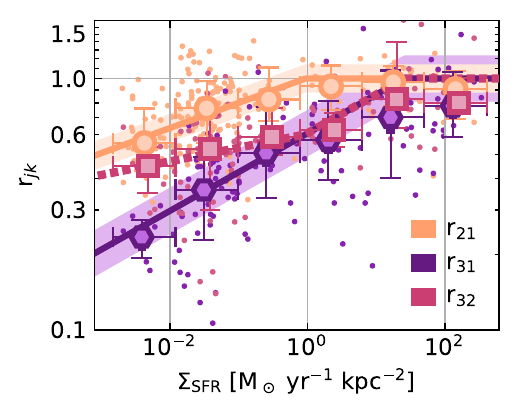}
    \caption{Summary of the relations between $r_{21}$ (orange), $r_{31}$ (purple), or $r_{32}$ (pink) with $\Sigma_{\rm SFR}$. Small markers show values from the combined literature and AMISS samples. Large markers show median values in six bins; vertical error bars show the 16th and 84th percentiles and horizontal error bars show the extent of each bin. Solid lines and filled regions are the prescriptions and intrinsic scatters derived from the AMISS data (Equations~\ref{eq:r21sigsfr} and \ref{eq:r31sigsfr}). The dashed pink line shows the ratio of these two prescriptions, which provides a good match to the $r_{32}$ data.}
    \label{fig:litsigma}
\end{figure}

\begin{figure*}
    \centering
    \includegraphics[width=\textwidth,trim={0 4mm 0 3mm},clip]{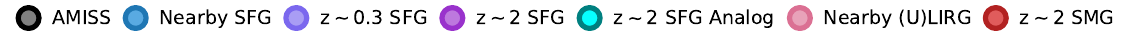}
    \includegraphics[width=\textwidth]{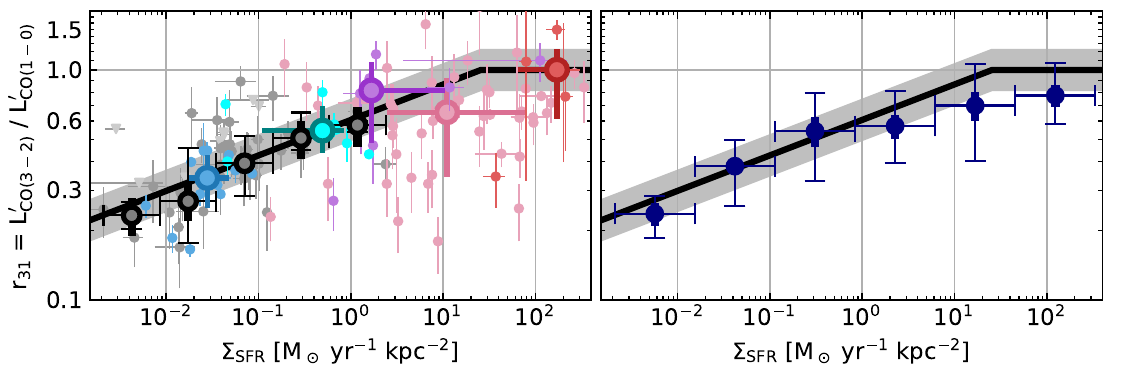}
    \includegraphics[width=\textwidth]{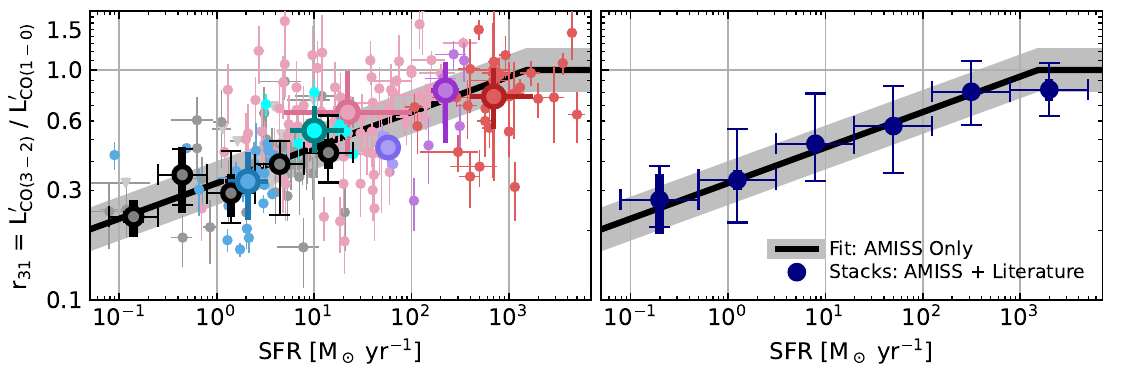}

    \caption{Left panels: $r_{31}$ from AMISS (gray) and literature (colored, see legend) plotted against $\Sigma_{\rm SFR}$ and SFR. Small, light colored points correspond to individual objects with error bars showing $1\sigma$ uncertainties (when reported). For literature samples, large points with a dark outline represent median and 16th-84th percentile ranges of $x$- and $r_{31}$-values; for the AMISS sample we show the median $r_{31}$ in five evenly spaced bins. The best fitting power law from the AMISS sample is shown by the black line, and the intrinsic scatter is shown by the gray filled region.
    Right panels: median values of all $r_{31}$ data, binned along the $x$-axis. Thin vertical bars show the 16th-84th percentile range of each bin; thick vertical bars show the uncertainty in the median. Horizontal bars show the $x$-range of each bin. We also reproduce our power law fits. Note that the while the black lines closely match all of the binned data, they are fitted using only the AMISS sample.}
    \label{fig:r31comp}
\end{figure*}
\begin{figure*}
    \centering
    \includegraphics[width=\textwidth,trim={0 4mm 0 3mm},clip]{f7legend.pdf}
    \includegraphics[width=\textwidth]{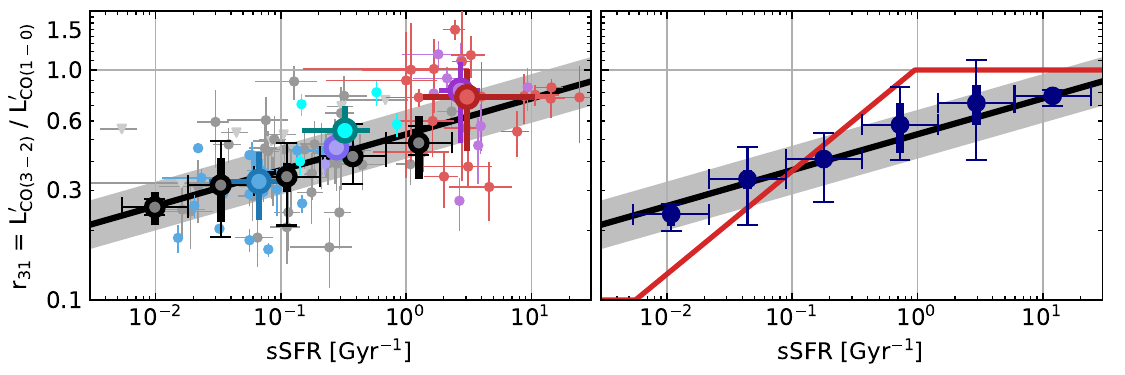}
    \includegraphics[width=\textwidth]{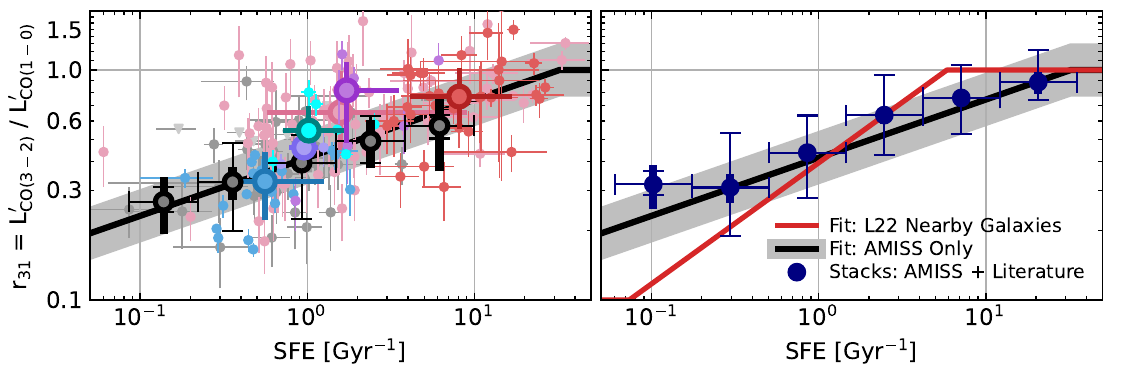}
    \caption{Similar to Figure~\ref{fig:r31comp}, but with sSFR and SFE as $x$-variables. In the right panel we show the fits proposed in \citet[]{leroy+22} as red lines.}
    \label{fig:r31comp2}
\end{figure*}
\begin{figure*}
    \centering
    \includegraphics[width=\textwidth,trim={0 4mm 0 3mm},clip]{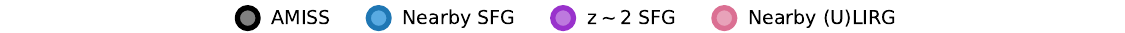}
    \includegraphics[width=\textwidth]{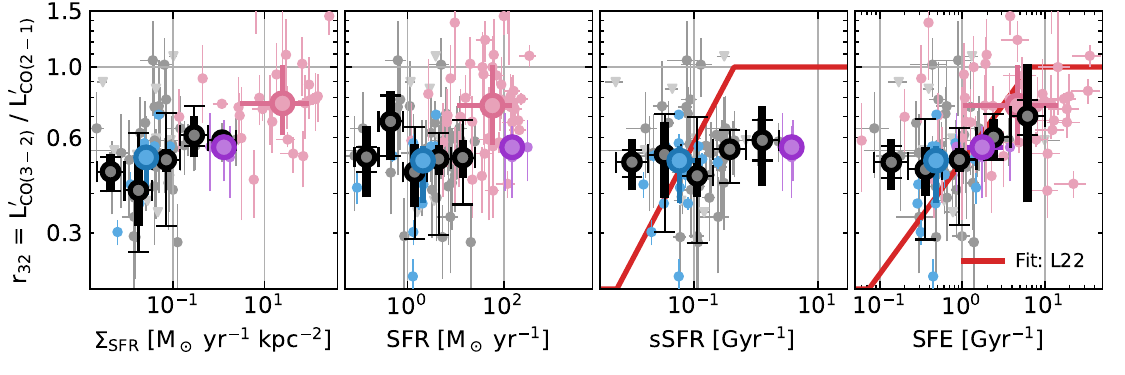}
    \caption{AMISS (gray) and literature (colored, see legend) $r_{32}$ values as a function of $\Sigma_{\rm SFR}$, SFR, sSFR, and SFE. Small, light colored points correspond to individual objects. For the literature samples large points with a dark outline represent the median and 16th-84th percentile range of $x$- and $r_{32}$- values within individual literature samples; for the AMISS sample we show the median $r_{32}$ in five evenly spaced bins of $x$. 
    Red lines show the proposed scaling relations from \citet{leroy+22}.}
    \label{fig:r32comp}
\end{figure*}

We have discussed the $r_{21}$ ratio extensively in \citetalias{keenan+24b}, and so focus here on comparing our $r_{31}$ and $r_{32}$ results to those found in the literature. To perform this comparison we have compiled line ratios measurements from nearby star forming galaxies \citep{leroy+22,lamperti+20}, 
IR selected galaxies \citep{papadopoulos+12,montoya-arroyave+23}, star forming galaxies at $z\sim0.3$ \citep{bauermeister+13a} and cosmic noon \citep[$1<z<3$;][]{daddi+15,bolatto+15,riechers+20}, cosmic noon-analogues \citep{lenkic+23}, and submillimeter selected high-redshift galaxies \citep{sharon+16,friascastillo+23,taylor+25}. Full details regarding the construction of our literature sample are given in Appendix~\ref{ap:lit}. Note that, due to differing availability of stellar mass and galaxy size information, not all objects/samples appear in all plots in this section.

Figure~\ref{fig:litsigma} summarizes how all three low-$J$ line ratios vary as a function of $\Sigma_{\rm SFR}$. After binning the data in $\Sigma_{\rm SFR}$, the $r_{21}$ and $r_{31}$ values of the combined sample clearly follow the prescriptions from Section~\ref{ss:presc}, even outside the parameter range used for the fits. The AMISS fits suggest that $r_{21}$ asymptotes one at $\Sigma_{\rm SFR}\gtrsim 1\,{\rm M_\odot\,yr^{-1}\,kpc^{-2}}$ while $r_{31}$ reaches the limit at $\Sigma_{\rm SFR}$ about 25 times higher. These limits are confirmed by the literature data which provides relatively good sampling of the high-$\Sigma_{\rm SFR}$ regime to complement the low- to mid-$\Sigma_{\rm SFR}$ AMISS data. In fact, fitting the parameters of Equations~\ref{eq:r21sigsfr} and~\ref{eq:r31sigsfr} using the binned points in Figure~\ref{fig:litsigma} would have given $\log r_{21}=0.11\log\Sigma_{\rm SFR}-0.01$ and $\log r_{31}=0.17\log\Sigma_{\rm SFR}-0.21$ (with maximum $r_{jk}$ fixed at unity) -- nearly identical to our fiducial prescriptions. There is some suggestion that the trends flatten as the ratios approach the unity, however this may also be an effect of combining a wide range of literature data with unknown calibration offsets and other systematics \citep[for a discussion of the difficulty comparing literature line ratios see][]{denbrok+21,keenan+24b}. The key takeaway from Figure~\ref{fig:litsigma} is the remarkably good performance of Equations~\ref{eq:r21sigsfr} and \ref{eq:r31sigsfr} in reproducing the observed trends in a wide variety of galaxies. 

More broadly, trends between $r_{31}$ and $\Sigma_{\rm SFR}$, SFR, sSFR, and/or SFE are reported in a number of prior works \citep{leroy+22,lamperti+20,montoya-arroyave+23,mao+10,yao+03,li+20,denbrok+23b}. In Figures~\ref{fig:r31comp} and \ref{fig:r31comp2} we show how literature and AMISS $r_{31}$ measurements correlate with these variables. While line ratios from the literature show a significant scatter, they generally follow similar trends to those found within the AMISS sample. Median values for bins and for individual literature surveys lie near our best-fitting power-laws (truncated at unity) for all four quantities. As with $\Sigma_{\rm SFR}$ in Figure~\ref{fig:litsigma}, the match of the literature data to the AMISS fits holds even when extrapolating beyond the range in SFR, sSFR, and SFE used for the fit.
In the upper left panel of Figure~\ref{fig:r31comp}, ULIRGs appear lie slightly below the trend followed by other samples in $\Sigma_{\rm SFR}$. This may be attributable partially to larger observational uncertainties for this sample -- as seen in the significant scatter -- and partially to the derivation of $\Sigma_{\rm SFR}$, which is based on far-infrared sizes for the highly dust-obscured ULIRGs, but optical sizes for most star forming galaxy samples.


Quantitatively, the trends we find between $r_{31}$ and sSFR or SFE are flatter than those proposed by \citet{leroy+22}, which we show in the right panels of Figure~\ref{fig:r31comp2} for comparison.
This can be attributed to the wider range in sSFR or SFE spanned by the AMISS sample. Both fits are in reasonable agreement with the data near $0.01\,{\rm Gyr}^{-1}\lesssim{\rm sSFR}\lesssim 0.3\, {\rm Gyr}^{-1}$ or $0.3\,{\rm Gyr}^{-1}\lesssim{\rm SFE}\lesssim 2\,{\rm Gyr}^{-1}$, where the \citet{leroy+22} sample resides. However when considering broader ranges in these parameters, our flatter relations are favored.

Figures~\ref{fig:litsigma}, \ref{fig:r31comp}, and \ref{fig:r31comp2} help to explain why studies of IR selected galaxies -- at both low and high redshifts -- often find no trend in CO line ratios \citep[e.g.][]{montoya-arroyave+23,friascastillo+23,sharon+16}. The ranges in $\Sigma_{\rm SFR}$, SFR, sSFR, and SFE probed in these studies typically lie in the regime where the line ratios approach the limit for optically thick gas, and minimal variation is to be expected. 

Only a handful of works have studied trends in $r_{32}$. \citet{leroy+22} found tentative evidence of a correlation between $r_{32}$ and sSFR, while \citet{montoya-arroyave+23} found that ULIRGs have higher $r_{32}$ than main sequence star forming galaxies.

We plot our compiled $r_{32}$ data as a function of $\Sigma_{\rm SFR}$, SFR, sSFR, and SFE in Figure~\ref{fig:r32comp}. The median $r_{32}$ for our sample is $0.52$, consistent with the value of 0.46 reported by \citet{leroy+22}, however our measurements rule out the relation between sSFR and $r_{32}$ proposed in therein. The AMISS results are consistent with a constant $r_{32}$ over $\sim2$ decades in sSFR as well as similar ranges in stellar mass, gas mass, SFR, and SFE.

The $r_{32}$ values of (U)LIRGs lie above those in star forming galaxies from the AMISS, \citet{leroy+22}, and our $z\sim2$ compilation. Warmer, denser, and more turbulent (lower opacity) ISM conditions in (U)LIRGs likely drive the higher $r_{32}$ values of those systems. SFR alone does not capture these conditions, and as such the (U)LIRGs are not cleanly separated from the star forming galaxies along this axis in Figure~\ref{fig:r32comp}. On the other hand, when treating $r_{32}$ as a function of $\Sigma_{\rm SFR}$ or SFE we observe a cleaner separation between (U)LIRGs and the other samples, as well as tentative trends among the main sequence galaxies.

In absence of noise we expect $r_{32}=r_{31}/r_{32}$; this is clear in Figure~\ref{fig:litsigma}, where the median $r_{32}$ values fall very close to the ratio of our $r_{31}$ and $r_{21}$ prescriptions. This means $r_{32}\sim r_{31}$ at $\Sigma_{\rm SFR}>1\,{\rm M_\odot\,yr^{-1}\,kpc^{-2}}$ (i.e. where $r_{21}\sim1$), but then falls off only gradually in the $\Sigma_{\rm SFR}$ range occupied by galaxies on the $z\sim0$ main sequence. This explains the weak-to-absent trends of $r_{32}$ in most of the samples, as well as the jump in $r_{32}$ for the (U)LIRG sample.


\subsection{Robustness of Trends}

AMISS primarily observed CO(3--2) targets with bright CO(1--0) and CO(2--1) lines. The resulting sample consists mostly of galaxies at the nearby end of the distance range for the parent sample or which have high SFRs and CO luminosities. As a result, our sample probes a wide range in star formation-related quantities (SFR, $\Sigma_{\rm SFR}$, SFR/SFR$_{\rm MS}$, sSFR, and SFE), but does not uniformly sample the stellar mass-SFR parameter space in Figure~\ref{fig:sample}.

To test whether the limitations of this sample bias our results, we can compare the $r_{21}$ fits in this paper with those from \citetalias{keenan+24b}, which used a larger, stellar mass-selected sample. The \citetalias{keenan+24b} sample had better coverage of the stellar mass-SFR plane (indicated by gray points in Figure~\ref{fig:sample}), with reliable $r_{21}$ measurements for all targeted galaxies with SFR above 0.5 M$_\odot$~yr$^{-1}$ and a significant fraction of galaxies with SFR above 0.1 M$_\odot$~yr$^{-1}$. Based on this larger sample, \citetalias{keenan+24b} concluded that $r_{21}$ is tightly correlated with SFR -- and many other star formation-related properties -- but shows no correlation with stellar mass. This suggests that adequate sampling along the SFR axis alone is sufficient to constrain trends in the low-$J$ CO line ratios, even with imperfect sampling in the stellar mass axis.

In Figures~\ref{fig:trends} and~\ref{fig:trends_summary} we show the \citetalias{keenan+24b} $r_{21}$ fits alongside our new results. For all choices of $x$-variable, the two $r_{21}$ fits are comparable within the uncertainties, indicating that our sample is not heavily biased with respect to the galaxy properties that determine CO line ratios. The excellent agreement of the literature samples -- each with a unique sample selection -- with the AMISS fits for $r_{31}$ in Section~\ref{ss:litcomp} provides further evidence that our results are robust against selection effects.

Still, the AMISS CO(3--2) sample may not be well suited for identifying secondary correlations of the line ratios with galaxy properties tied to stellar mass. An expanded sample of CO(3--2) observations could strengthen future research on the $r_{31}$ and $r_{32}$ ratios, and may be especially valuable for $r_{31}$ owing to its larger dynamic range relative $r_{21}$ and $r_{32}$.

\section{The Physical Conditions of Molecular Gas in Star Forming Galaxies}\label{sec:ism_conditions}

\begin{figure*}
    \centering
    \includegraphics[width=\textwidth]{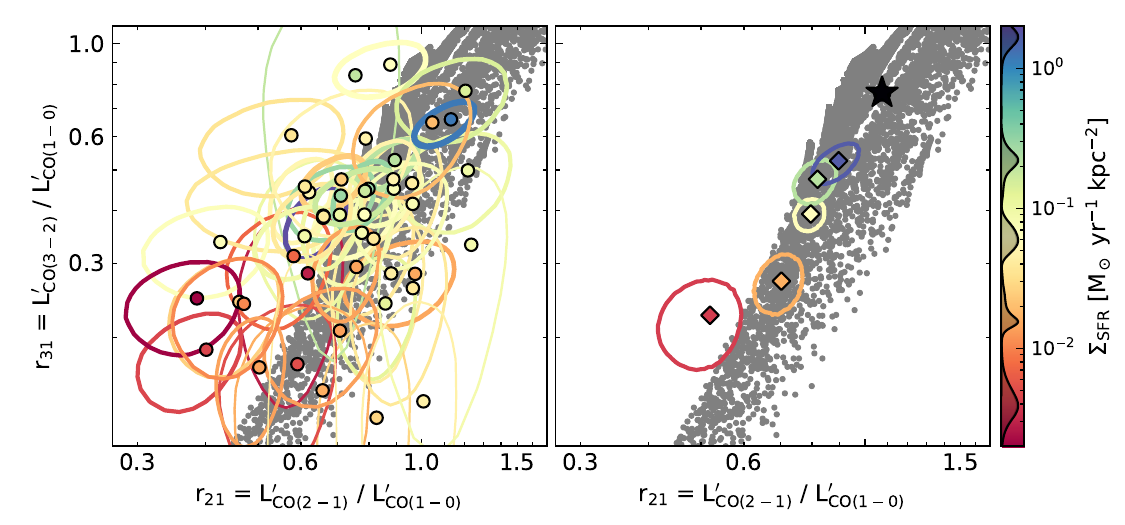}
    \caption{Left: $r_{21}$ and $r_{31}$ ratios of our sample. Points and their surrounding ellipses show the measured value and $1\sigma$ error region for ratio pairs from each galaxy. Galaxies are color coded by $\Sigma_{\rm SFR}$. The thickness of the error contour corresponds to the SNR of the CO(3--2) line, with the thinnest lines corresponding to galaxies where the CO(3--2) line is formally undetected. Gray points in the background correspond to the grid of cloud models from \citet{leroy+22}, showing the expected range of line ratios for individual clouds with a log-normal density distributions and constant temperatures. Right: median $r_{21}$ and $r_{31}$ in five bins of $\Sigma_{\rm SFR}$. Confidence contours ($1\sigma$) are determined by repeatedly perturbing each luminosity and $\Sigma_{\rm SFR}$ according to its uncertainty and rebinning the data. Black curves enclosing gray filled regions on the color bar show the distribution of the median $\Sigma_{\rm SFR}$ in each bin for the draws. The dark colored star shows the median line ratios for a collection of ULIRGs ($\Sigma_{\rm SFR}\sim10^{1.5}\,{\rm M_\odot\,yr^{-1}\,kpc^{-2}}$) from \citet{montoya-arroyave+23}.}
    \label{fig:rr}
\end{figure*}

\begin{figure*}
    \centering
    \includegraphics[width=\textwidth]{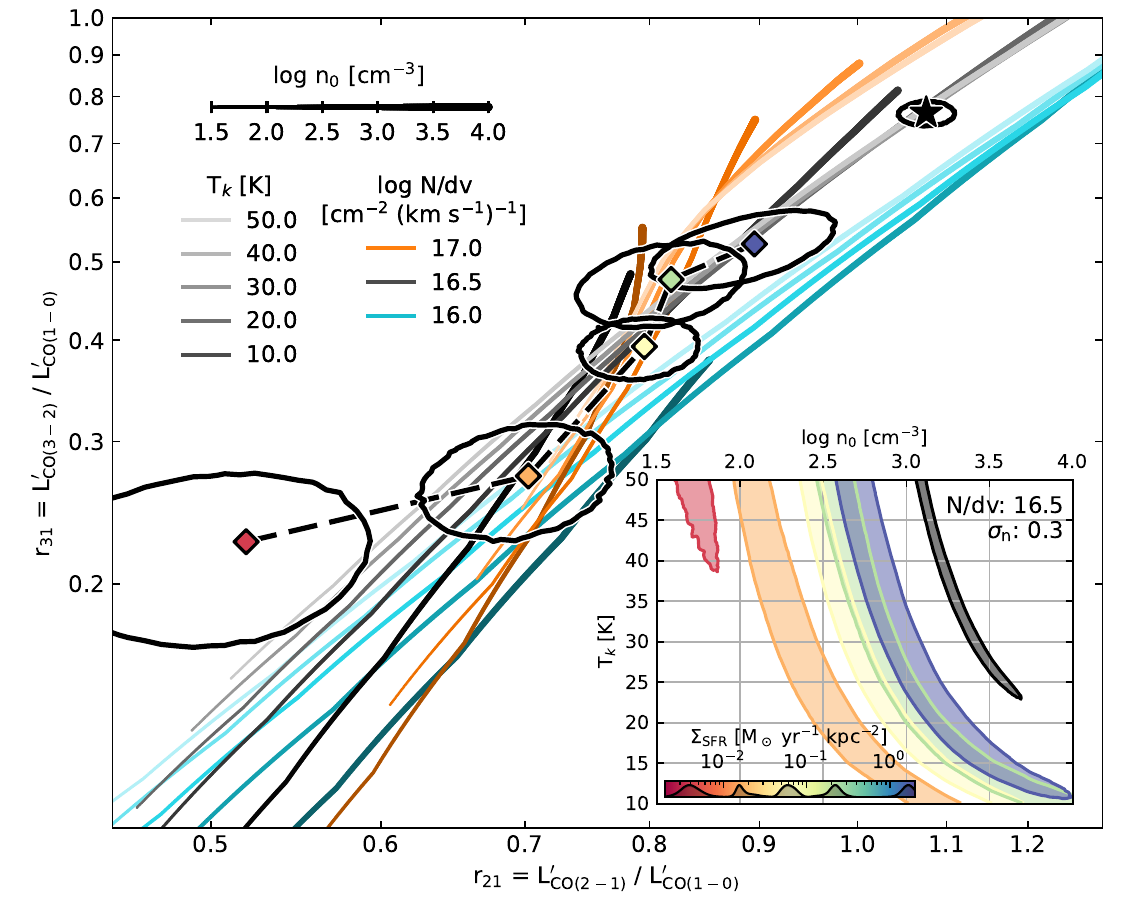}
    \caption{Main figure: solid lines shows the track in $r_{21}$ and $r_{32}$ when the mean density ($n_0$) of a \citet{leroy+22} molecular cloud model is varied, holding other parameters ($\sigma_n$, $T_k$, and $N/dv$) constant. The thickness of the lines indicates $n_0$ at that point along the track. Each track is color coded according to its $N/dv$, while the hue of the line indicates $T_k$. In general models move up and to the right a density increases. We overplot the median line ratios and their error regions for our $\Sigma_{\rm SFR}$ bins (diamonds) and for ULIRGs (black star) from the right panel of Figure~\ref{fig:rr}. As $\Sigma_{\rm SFR}$ increases, galaxy averaged line ratios roughly follow the tracks of increasing density at $N/dv\sim10^{16.5}\,{\rm cm^{-2}\,(km\,s^{-1})^{-1}}$.
    Inset: Filled contours represent the range of $n_0$ and $T_k$ for models that produce $r_{21}$ and $r_{31}$ values within $1\sigma$ of our binned data (i.e. lying within the error ellipses in the main figure). Here we show only models with $N/dv=10^{16.5}\,{\rm cm^{-2}\,(km\,s^{-1})^{-1}}$ and $\sigma_n=0.3$~dex. Similar plots for other combinations of $N/dv$ and $\sigma_n$ are shown in Appendix~\ref{ap:morefits}.}
    \label{fig:rrmodels}
\end{figure*}

At the scale of individual molecular clouds, CO line ratios are set by the gas density and temperature profiles, the spatial and velocity profiles of optical depth in each transition, and the CO-to-H$_2$ abundance ratio. Observed line ratios provide constraints on these properties, and, if degeneracies between the physical parameters can be sufficiently broken, may allow us to determine the physical state of the ISM. To accomplish this, we compare line ratios from our sample to the grid of cloud models from \citet{leroy+22}. These models consist of clouds with a lognormal H$_2$ density distribution characterized by a mean density $n_0$ and width $\sigma_n$, uniform kinetic temperature $T_k$, and uniform CO column density per line width $N/dv$. CO intensities and line ratios were computed using RADEX \citep{vandertak+07} to evaluate the non-LTE radiative transfer on a grid of one-zone, expanding sphere models (i.e. single $n_{\rm H_2}$, $T_k$, and $N/dv$), then summing the emission from models of differing $n_{\rm H_2}$ according to the density distribution of each cloud \citep{leroy+17,leroy+22}. The full model grid spans $T_k$ of 10~K to 50~K, $n_0$ of $10^{1.5}\,{\rm cm^{-3}}$ to $10^{5}\,{\rm cm^{-3}}$, $\sigma_n$ of 0.2~dex to 1.2~dex, and $N/dv$ of $10^{15}\,{\rm cm^{-2}\,(km\,s^{-1})^{-1}}$ to $10^{18}\,{\rm cm^{-2}\,(km\,s^{-1})^{-1}}$.

Extracting physical conditions for individual galaxies or a single bin by comparison to the model grid is infeasible; in most cases, a given set of line ratios can be reproduced by a multitude of models spanning a wide range of physical conditions. However, if we assume that cloud physical conditions change in a continuous manner in response to changes in $\Sigma_{\rm SFR}$, we can seek to match the observed $\Sigma_{\rm SFR}$-line ratio trends through variations of model parameters. The small intrinsic scatter in line ratios at fixed $\Sigma_{\rm SFR}$ suggests that this approach is reasonable; if ISM properties were only loosely coupled to $\Sigma_{\rm SFR}$, we would expect to see a larger dispersion in the line ratios. The unresolved nature of our observations requires the further assumption that cloud physical properties average in the same manner as line ratios over full galaxies. This is reasonable if variations in cloud properties across the galaxy are small or if a single cloud type typically dominates the CO emission. In a study of nine nearby galaxies resolved at kiloparsec scales, \citet{denbrok+21} find that line ratios are relatively constant over galaxy disks, with an intrinsic scatter of $\sim10\%$, suggesting this is a good first approximation \citep[see also][]{leroy+22}. In detail, some regions of galaxies -- especially centers -- are known to have elevated line ratios, which may alter the global average \citep{sawada+01,denbrok+23b,anirudh+25} Resolved line ratio measurements for large galaxy samples will be necessary to clarify the impact of these regions.

In Figure~\ref{fig:rr} we plot the $r_{21}$ and $r_{31}$ values of our sample and the model grid. Among our observations, ratio-pairs show a large scatter -- a result of the large statistical uncertainties for individual measurements. Nonetheless, a gradient is apparent as a function of $\Sigma_{\rm SFR}$ (shown by the color-coding of the observational data). This trend is clear when considering the median line ratios in bins of $\Sigma_{\rm SFR}$ (right panel). To show how the trend extends to more extreme $\Sigma_{\rm SFR}$, we also include median ratios for 11 ULIRGs from \citet{montoya-arroyave+23} in the right panel, denoted by a large star. We do not have $\Sigma_{\rm SFR}$ values for all galaxies in this ULIRG bin, but the median value for those we do have is $43\,{\rm M_\odot\,yr^{-1}\,kpc^{-2}}$, in line with typical (U)LIRG $\Sigma_{\rm SFR}$ values ranging from $\sim 1-100\,{\rm M_\odot\,yr^{-1}\,kpc^{-2}}$. This places these objects well above the range in $\Sigma_{\rm SFR}$ probed by our sample.

To illustrate how physical conditions translate to observed line ratios, the main panel of Figure~\ref{fig:rrmodels} shows tracks in the $r_{21}$--$r_{31}$ plane produced by varying $T_k$, $n_0$ and $N/dv$ in the model clouds. For simplicity, we fix $\sigma_n=0.3$~dex in these tracks based on the findings of \citet{denbrok+23b} for NGC~3627. Increasing $\sigma_n$ broadens the density distribution, resulting in more gas at high density for a given mean density. This qualitatively produces similar changes in line ratios to increasing $n_0$. 

The range of line ratios seen in our sample can be reproduced by variations in $n_0$ and $T_k$ in gas with $\sigma_n\sim0.3$~dex and $N/dv$ between $10^{16.5}$ and $10^{17}\,{\rm cm^{-2}\,(km\,s^{-1})^{-1}}$. Comparison between our $\Sigma_{\rm SFR}$ bins and the model tracks in Figure~\ref{fig:rrmodels} is consistent with an overall picture in which increasing fractions of molecular gas at high density drive galaxies to higher star formation efficiencies and $\Sigma_{\rm SFR}$ \citep{gao+04,kennicutt+21,narayanan+14,saintonge+17}. Meanwhile, feedback from star formation injects energy into the clouds increasing $T_k$ and possibly lowering the gas opacity. These effects have previously been noted in ULIRGs \citep{papadopoulos+12,kamenetzky+14}; our results connect this picture to galaxies near the main sequence, showing a relatively continuous evolution in line ratios and gas conditions is possible over a large range of $\Sigma_{\rm SFR}$. 

In the inset of Figure~\ref{fig:rrmodels}, we fix $\sigma_n=0.3$~dex and $N/dv=10^{16.5}\,{\rm cm^{-2}\,(km\,s^{-1})^{-1}}$ and show the range of $T_k$ and $n_0$ for models that lie within the $1\sigma$ confidence ellipses for each $\Sigma_{\rm SFR}$ bin (here, we limit our discussion here primarily to $\sigma_n=0.3$~dex and $N/dv=10^{16.5}\,{\rm cm^{-2}\,(km\,s^{-1})^{-1}}$, as other parameter choices tend to favor unexpectedly low densities or high temperatures; Figure~\ref{fig:morefits} in Appendix~\ref{ap:morefits} shows similar plots for a range of $\sigma_n$ and $N/dv$).  Considering first the bins with $0.01\,{\rm M_\odot yr^{-1} kpc^{-2}}<\Sigma_{\rm SFR}<2\,{\rm M_\odot yr^{-1} kpc^{-2}}$, we find that $T_k$ and $n_0$ are highly degenerate among the models that can match the bin-median line ratios. The observed trends in $r_{jk}$ with $\Sigma_{\rm SFR}$ can be explained by a factor of 10 increase in mean gas density at a fixed temperature or a rise in temperature from 10~K to 30~K at a fixed $n_0\sim10^{3}\,{\rm cm^{-3}}$. Explaining correlations between $r_{jk}$ and $\Sigma_{\rm SFR}$ as solely a temperature effect would require that $\Sigma_{\rm SFR}$ trace small changes in gas temperature with almost no scatter introduced by density variations or other effects. Some combination of temperature and density variations is most likely at play. The overall sense of these variations is that gas increases (or at least remains constant) in density as $\Sigma_{\rm SFR}$ rises from $0.01{\rm\,M_\odot\,yr^{-1}\,kpc^{-2}}$ to $2{\rm\,M_\odot\,yr^{-1}\,kpc^{-2}}$, while changes in temperature are less constrained.

Turning next to the ULIRG bin, we see that these extreme starbursts must have either denser or warmer (or both) cloud conditions than the lower $\Sigma_{\rm SFR}$ systems. Given the more turbulent ISM in ULIRGs, broader gas density distributions are plausible for these systems. Increasing either the mean density $n_0$ and the width of the density distribution $\sigma_n$ can place in more CO in dense parts of the cloud, resulting in similar increases in the CO line ratios. The central column of Figure~\ref{fig:morefits} shows that the change in line ratios from the $\Sigma_{\rm SFR}\sim 2{\rm\,M_\odot\,yr^{-1}\,kpc^{-2}}$ bin to the ULIRG bin could be explained by a factor of $\sim10$ increase in $n_0$ or an increase of $\sigma_n$ from 0.3~dex to 0.6~dex at fixed $n_0$. In either case, the range of allowed $T_k$ and $n_0$ are consistent with the those found by \citet{kamenetzky+14}, who used a two-zone cloud model to conclude that the low-$J$ CO SLEDs of ULIRGs are dominated by gas with densities between $10^{2.5}\,{\rm cm^{-3}}$ and $10^{3.5}\,{\rm cm^{-3}}$ and temperatures around 10 to 100~K.  Prior studies have suggested that ULIRGs have dramatically different molecular cloud conditions to other star forming systems, primarily by comparing them to clouds in the Milky Way \citep{papadopoulos+12}. Our results instead highlight that star forming galaxies exhibit a wide range of cloud conditions. ULIRGs lie at one extreme of a continuum of conditions, however similar conditions can also be seen locally within other classes of galaxies \citep{sun+20,denbrok+23b}. Figure~\ref{fig:rrmodels} shows that the changes in physical conditions between many of the AMISS $\Sigma_{\rm SFR}$ bins are similar in magnitude to the change between the $\Sigma_{\rm SFR}\sim 2{\rm\,M_\odot\,yr^{-1}\,kpc^{-2}}$ galaxies and ULIRGs.

Finally, our lowest $\Sigma_{\rm SFR}$ bin lies on the edge of the $r_{21}$--$r_{32}$ space covered in the model grid, and requires gas with high temperature, low density, and $N/dv\sim10^{16.5}\,{\rm cm^{-2}\,(km\,s^{-1})^{-1}}$. The low density is consistent with a significant fraction of the molecular gas in these quiescent systems being non-star forming. 

In practice, galaxy-integrated line ratios arise from an ensemble of clouds that likely have complex distributions of density, temperature, and optical depth not fully captured by single-cloud models. While our use of realistic density distributions represents a significant improvement over single- or two-zone models that are often used to fit SLEDs, the conclusions we draw will require further refinement using even more sophisticated modeling. Constraining such models is not possible with only the unresolved, low-$J$ CO SLED. However, a number of promising paths forward exist, including the use of additional molecular species and/or dust \citep{harrington+21}, adding information from mid- to high-$J$ CO transitions \citep{liu+21}, and fitting models directly at the cloud scale using high resolution observations of nearby galaxies \citep{denbrok+23b}.


\section{Conclusion}\label{sec:conclusion}

We have presented the low-$J$ CO spectral line energy distributions and line ratios for a sample of 47 star forming galaxies at $z\sim0$. The galaxies within our sample exhibit diverse SLEDs, with galaxies of higher SFR, $\Sigma_{\rm SFR}$, and $\Sigma_{\rm mol}$ lying closer to expectations for optically thick gas in local thermal equilibrium. These trends are qualitatively consistent with theoretical expectations, but the range of line ratios is larger than that seen in commonly used simulation-based prescriptions \citep[e.g.][]{narayanan+14}. 

We find that the $r_{21}$ and $r_{31}$ line ratios are positively correlated with $\Sigma_{\rm SFR}$, $\Sigma_{\rm mol}$, SFR, sSFR, and SFE, but not with stellar mass. While many prior studies have found weak trends in $r_{31}$ with other galaxy properties or pointed to differences in line ratios between starburst and main-sequence galaxies, our sample spans a wide enough range in $\Sigma_{\rm SFR}$ and other key properties to clearly reveal the smooth rise in $r_{31}$ over a factor of 300 variation in $\Sigma_{\rm SFR}$ or SFR. With the inclusion of starbursts and high-$z$ galaxies from the literature, we show that these trends extend over four to five orders of magnitude.
    
Knowledge of the low-$J$ CO SLED and line ratio is useful for converting between different CO lines when measuring molecular gas masses. We give empirical prescriptions for estimating $r_{j1}$ based on either SFR or $\Sigma_{\rm SFR}$ in Equations~\ref{eq:r21prescription} through~\ref{eq:r31sigsfr}. These prescriptions capture the full range in variations of $r_{j1}$ for both our sample and compiled literature samples, and appear to be reliable for both local and high redshift galaxies. 

While both the SFR and $\Sigma_{\rm SFR}$ scaling relations reproduce the observed $r_{21}$ and $r_{31}$ values, theory suggests that the physical conditions of clouds which give rise to the line ratios -- distributions of gas temperature, H$_2$ density, and optical depth -- should be more closely tied to surface density-like quantities. This is consistent with our finding tentative evidence that $r_{32}$ correlates with $\Sigma_{\rm SFR}$ and $\Sigma_{\rm mol}$ but not SFR or $M_{\rm mol}$.

By investigating which variations in physical properties of molecular cloud models can match the observed trends of $r_{21}$ and $r_{31}$ with $\Sigma_{\rm SFR}$, we can gain insight into how ISM conditions differ across the galaxy population. This population-based approach somewhat alleviates the parameter degeneracies that complicate extracting physical conditions of individual galaxies using low-$J$ CO line ratios. ISM conditions in main sequence and starburst galaxies lie along a continuum in molecular cloud conditions. Along this continuum, progressively increasing gas density gives rise to higher SFR density and $\Sigma_{\rm SFR}$. Increasing star formation may in turn increase the radiation field (and other forms of feedback) to which the molecular gas is exposed, further affecting the gas temperature. This series of effects produces the significant correlations of CO line ratios with $\Sigma_{\rm SFR}$ and $\Sigma_{\rm mol}$. Within the parameter space covered by the \citet{leroy+22} molecular cloud models, we find mean H$_2$ densities ranging from $<10^2\,{\rm cm^{-3}}$ at $\Sigma_{\rm SFR}=10^{-2.5}\,{\rm M_\odot\,yr^{-1}\,kpc^{-2}}$ to $>10^3 \,{\rm cm^{-3}}$ in ULIRGs with $\Sigma_{\rm SFR}\sim10^{1.5}\,{\rm M_\odot\,yr^{-1}\,kpc^{-2}}$. For starburst systems, somewhat lower mean densities with wider density distributions are also possible. This suggests higher $\Sigma_{\rm SFR}$ corresponds to a higher fraction of gas above the density threshold for star formation, consistent with the idea that starburst galaxies have higher star formation efficiencies. On the other hand, the molecular gas in quiescent galaxies appears to be warm ($T_k>40$~K) and low density, suggesting the bulk of molecular clouds in these systems lie below the density thresholds for gravitation collapse and star formation.

The low-$J$ CO SLEDs of star forming galaxies contain significant information about ISM conditions. However, constraining detailed models of the molecular ISM in individual galaxies using only these data is very difficult. Here we have shown what insights can be gained from a population-based approach with a large, diverse sample of galaxies with high quality measurements of CO(1--0) through CO(3--2). Expanding the sample of CO(3--2) observations to better match the CO(1--0) and CO(2--1) coverage of main-sequence and quiescent galaxies could further improve our population constraints. Further progress can be made by incorporating additional species such as CO isotopologues, dense gas tracers, and dust. 
These data can break degeneracies between density, temperature, and optical depth and constrain more sophisticated ISM models. A number of recent works have demonstrated the promise of various parts of this approach \citep{leroy+17,harrington+21,liu+21}. 
At the same time, highly resolved observations of nearby galaxies, which have become routine with ALMA, can further clarify the physical picture of the ISM and help put better priors on molecular cloud models \citep{denbrok+23b,egusa+22,maeda+23,deng+25,komugi+25,koda+25,denbrok+25}.

\acknowledgments

The authors thank the Arizona Radio Observatory operators, engineering and management staff, without whom this research would not have been possible. We also thank L. A. Boogaard and F. Walter for their feedback on many aspects of this manuscript and A. K. Leroy and T. D\'iaz-Santos for sharing some of the data in our literature line ratio compilation. We thank the anonymous referee, whose detailed feedback helped clarify the presentation and discussion of our results. RPK thanks J. R. Keenan for her tireless support and input on many key figures. DPM and RPK were supported in part by the National Science Foundation through grant AST-2308041. 

This paper makes use of data collected by the the UArizona ARO Submillimeter Telescope and the UArizona ARO 12-meter Telescope, the IRAM 30m telescope, and the Sloan Digital Sky Survey. The UArizona ARO 12-meter Telescope on Kitt Peak and the UArizona ARO Submillimeter Telescope on Mt. Graham are operated by the Arizona Radio Observatory (ARO), Steward Observatory, University of Arizona. IRAM is supported by INSU/CNRS (France), MPG (Germany) and IGN (Spain). Funding for the SDSS and SDSS-II was provided by the Alfred P. Sloan Foundation, the Participating Institutions, the National Science Foundation, the U.S. Department of Energy, the National Aeronautics and Space Administration, the Japanese Monbukagakusho, the Max Planck Society, and the Higher Education Funding Council for England. The SDSS website is \url{www.sdss.org}.


\appendix
\section{Power Law Fit Results}\label{ap:fit_results}

\begin{deluxetable*}{l|l|rr|rl}
    \tablecaption{Regression parameters: $\log r_{jk} = m \log x + b + \epsilon(s)$ \label{tab:fits}}
    \tablehead{
        \colhead{$x$} & \colhead{Fit} & \colhead{$\rho_{mb}$} & \colhead{$m/\sigma_m$} & \multicolumn{2}{c}{$r$ ($p$)}\\
        \colhead{(1)} & \colhead{(2)} & \colhead{(3)} & \colhead{(4)} & \multicolumn{2}{c}{(5)}
    }
    \startdata
        \hline
        \multirow{3}{*}{$M_*$ [$10^{10}$ M$_\odot$]} & $\log r_{21} = (0.08\pm0.04)\log x - (0.16\pm0.02),\, s = 0.08\pm0.02$ & $-0.73$ & $1.8$ & 0.32 & \hspace{-.7em}(2.8\%) \\
            & $\log r_{31} = (0.08\pm0.07)\log x - (0.44\pm0.04),\, s = 0.13\pm0.03$ & $-0.68$ & $1.1$ & 0.15 & \hspace{-.7em}(36.8\%) \\
            & $\log r_{32} = (-0.03\pm0.05)\log x - (0.26\pm0.03),\, s \leq 0.14$ & $-0.71$ & $-0.6$ & -0.25 & \hspace{-.7em}(14.6\%) \\
        \hline
        \multirow{3}{*}{SFR [M$_\odot$ yr$^{-1}$]} & $\log r_{21} = (0.14\pm0.03)\log x - (0.21\pm0.02),\, s = 0.04\pm0.02$ & $-0.76$ & $5.0$ & 0.62 & \hspace{-.7em}($<$0.1\%) \\
            & $\log r_{31} = (0.15\pm0.04)\log x - (0.49\pm0.03),\, s = 0.10\pm0.03$ & $-0.67$ & $3.5$ & 0.35 & \hspace{-.7em}(3.9\%) \\
            & $\log r_{32} = (0.00\pm0.04)\log x - (0.27\pm0.03),\, s \leq 0.14$ & $-0.76$ & $0.1$ & -0.23 & \hspace{-.7em}(16.9\%) \\
        \hline
        \multirow{3}{*}{sSFR [Gyr$^{-1}$]} & $\log r_{21} = (0.10\pm0.03)\log x - (0.04\pm0.03),\, s = 0.06\pm0.02$ & $0.87$ & $3.3$ & 0.44 & \hspace{-.7em}(0.2\%) \\
            & $\log r_{31} = (0.16\pm0.05)\log x - (0.28\pm0.05),\, s = 0.11\pm0.03$ & $0.88$ & $3.0$ & 0.29 & \hspace{-.7em}(8.4\%) \\
            & $\log r_{32} = (0.02\pm0.04)\log x - (0.25\pm0.04),\, s \leq 0.14$ & $0.85$ & $0.6$ & -0.07 & \hspace{-.7em}(68.3\%) \\
        \hline
        \multirow{3}{*}{SFR / SFR$_{\rm MS}$} & $\log r_{21} = (0.14\pm0.03)\log x - (0.19\pm0.02),\, s = 0.05\pm0.02$ & $-0.69$ & $4.6$ & 0.57 & \hspace{-.7em}($<$0.1\%) \\
            & $\log r_{31} = (0.19\pm0.05)\log x - (0.48\pm0.03),\, s = 0.10\pm0.03$ & $-0.62$ & $3.7$ & 0.35 & \hspace{-.7em}(3.9\%) \\
            & $\log r_{32} = (0.02\pm0.04)\log x - (0.28\pm0.03),\, s \leq 0.14$ & $-0.73$ & $0.4$ & -0.17 & \hspace{-.7em}(32.8\%) \\
        \hline
        \multirow{3}{*}{$\Sigma_{\rm SFR}$ [M$_\odot$ yr$^{-1}$ kpc$^{-2}$]} & $\log r_{21} = (0.08\pm0.02)\log x - (0.02\pm0.03),\, s = 0.06\pm0.02$ & $0.90$ & $3.4$ & 0.47 & \hspace{-.7em}($<$0.1\%) \\
            & $\log r_{31} = (0.16\pm0.04)\log x - (0.21\pm0.05),\, s = 0.09\pm0.03$ & $0.90$ & $4.2$ & 0.51 & \hspace{-.7em}(0.1\%) \\
            & $\log r_{32} = (0.05\pm0.03)\log x - (0.21\pm0.04),\, s \leq 0.13$ & $0.87$ & $1.7$ & 0.17 & \hspace{-.7em}(32.1\%) \\
        \hline
        \multirow{3}{*}{$M_{\rm mol}$ [$10^{9}$ M$_\odot$]} & $\log r_{21} = (0.11\pm0.04)\log x - (0.20\pm0.03),\, s = 0.07\pm0.02$ & $-0.86$ & $2.8$ & 0.38 & \hspace{-.7em}(0.8\%) \\
            & $\log r_{31} = (0.13\pm0.06)\log x - (0.49\pm0.04),\, s = 0.12\pm0.03$ & $-0.80$ & $2.2$ & 0.18 & \hspace{-.7em}(28.4\%) \\
            & $\log r_{32} = (0.01\pm0.04)\log x - (0.28\pm0.04),\, s \leq 0.14$ & $-0.85$ & $0.1$ & -0.26 & \hspace{-.7em}(12.2\%) \\
        \hline
        \multirow{3}{*}{$\Sigma_{\rm mol}$ [$10^{8}$ M$_\odot$ kpc$^{-2}$]} & $\log r_{21} = (0.07\pm0.03)\log x - (0.12\pm0.02),\, s = 0.07\pm0.02$ & $0.29$ & $2.1$ & 0.22 & \hspace{-.7em}(13.3\%) \\
            & $\log r_{31} = (0.19\pm0.05)\log x - (0.38\pm0.02),\, s = 0.10\pm0.03$ & $0.36$ & $3.9$ & 0.42 & \hspace{-.7em}(1.0\%) \\
            & $\log r_{32} = (0.08\pm0.04)\log x - (0.27\pm0.02),\, s \leq 0.13$ & $0.08$ & $2.2$ & 0.23 & \hspace{-.7em}(18.1\%) \\
        \hline
        \multirow{3}{*}{SFE [Gyr$^{-1}$]} & $\log r_{21} = (0.16\pm0.05)\log x - (0.10\pm0.02),\, s = 0.06\pm0.02$ & $0.40$ & $3.6$ & 0.48 & \hspace{-.7em}($<$0.1\%) \\
            & $\log r_{31} = (0.25\pm0.09)\log x - (0.37\pm0.03),\, s = 0.12\pm0.03$ & $0.58$ & $2.8$ & 0.35 & \hspace{-.7em}(3.6\%) \\
            & $\log r_{32} = (0.00\pm0.07)\log x - (0.27\pm0.02),\, s \leq 0.14$ & $0.34$ & $0.0$ & -0.02 & \hspace{-.7em}(90.2\%) \\
    \enddata
    \tablecomments{Columns are (1) $x$-variable used in fit; (2) power-law fit results for $r_{21}$, $r_{31}$, and $r_{32}$ as a function of $x$; (3) the correlation coefficient between uncertainties in $m$ and $b$ ($\rho_{mb}\sigma_m\sigma_b$ gives the off-diagonal term in the covariance matrix for $m$ and $b$); (4) the ratio between the fitted slope and its uncertainty; (5) the Pearson correlation coefficient and $p$-value between $x$ and $r_{jk}$. Upper limits on $s$ in the $r_{32}$ fits are given at the 95\% confidence level.}
\end{deluxetable*}

In Table~\ref{tab:fits} we provide the best fitting parameters and uncertainties for the power law fits described in Section~\ref{sec:ratios}, the degrees by which the slopes deviate from zero ($m/\sigma_m$), and correlation coefficients for uncertainties on the $m$ and $b$ fit parameters ($\rho_{mb}$). The covariance of uncertainties in the best fitting $m$ and $b$ values is ${\rm cov}(m,b) = \rho_{mb}\sigma_m\sigma_b$. 

We also report Pearson correlation coefficients $r$ and the $p$-values for a null hypothesis of no correlation. We recommend the significance of the best-fitting slope, $m/\sigma_m$, as a better indicator of our confidence in the correlations between the $x$-variables and $r_{jk}$, as our sample has a non-Gaussian distribution in many of the $x$-variables considered and individual measurements can have large uncertainties which are not accounted for in the standard computation of $r$.

\section{Data Included in our Literature Compilation}\label{ap:lit}

In this Appendix we provide details and references for the sources entering our literature compilation in Section~\ref{ss:litcomp}. Whenever possible, we tabulate line ratios, CO(1--0) luminosities, SFRs, stellar masses, and half-light radii for each literature source. When SFRs are not reported, we derive them from IR luminosity, preferring IR luminosities with AGN contributions removed when provided. We derive molecular gas masses from the reported CO(1--0) luminosities, using a starburst $\alpha_{\rm CO}=1.0$~M$_\odot$\,(K\,km\, s$^{-1}$\,pc$^{-2}$)$^{-1}$ for SMGs and ULIRGs and \citep{downes+98} and a Milky Way-like value $\alpha_{\rm CO}=4.3$~M$_\odot$\,(K\,km\, s$^{-1}$\,pc$^{-2}$)$^{-1}$ for the remaining sources \citep{bolatto+13,daddi+10}. If we instead adopted a constant conversion factor for all samples \citep[see e.g.][]{dunne+22}, the starbursts would shift to the left in the SFE panels of Figure~\ref{fig:r31comp2} and~\ref{fig:r32comp}, moving them off of the AMISS trend; other results would remain unchanged. We derive star formation and molecular gas surface densities as half of the global SFR or $M_{\rm mol}$ divided by the area within the half-light radius of the galaxy.

\subsection{Star Forming Galaxies}

\citet{leroy+22} provide galaxy-integrated line ratio measurements derived from single-dish mapping of nearby galaxies. We include $r_{31}$ values for 19 galaxies and $r_{32}$ for 27 galaxies from this sample. When multiple measurements for a single galaxy are present, we prefer those using CO(3--2) data from APEX, CO(2--1) data from ALMA, and CO(1--0) data from the Nobeyama 45-m. We take CO luminosities, line ratios, stellar masses, and SFRs from their Table 2; for the subset of galaxies covered by the PHANGS-ALMA survey we take half-light radii from Table 4 of \citet{leroy+21}. 

\citet{lamperti+20} include $r_{31}$ measurements of three additional xCOLD GASS galaxies not included in our main sample because they were not observed by the SMT in CO(2--1). We include the additional $r_{31}$ measurements from this dataset, with line ratios and ancillary data processed in the same manner described in Section~\ref{sec:survey}.

\citet{bauermeister+13a} provide $r_{31}$ measurements for four $z\sim0.3$ star forming galaxies observed with CARMA. We use their galaxy-mean $r_{31}$ values (Table 4), excluding one galaxy undetected in CO(3--2). We take SFR and stellar mass information from their Table 1 and CO(1--0) luminosities from their Table 2.

\citet{daddi+15} report $r_{31}$ for three galaxies and $r_{32}$ for four galaxies at $z\sim1.5$ based on observations with the VLA and IRAM PdBI. We use the CO fluxes and line ratios reported in their Table 2 and stellar masses and IR luminosities in their Table 3, and retrieve half-light radii from Tabel 3 of \citet{daddi+10}. 

\citet{bolatto+13} provide $r_{31}$ for two star forming galaxies at $z\sim2$, also based on observations with the VLA and PdBI. We use the CO luminosities derived from the C+D configurations reported in their Table 1 and line ratios and ancillary information reported in their Table 2. We take the half-light radii to be 4.6 kpc for BX610 and 4.8 kpc for MD94 based on the discussion in their Section~3.2.

\citet{riechers+20} measure $r_{31}$ for CO-selected galaxies at $z\sim2.5$ using ALMA and the VLA. We include the three sources with high-confidence $r_{31}$ values (9mm.1, 9mm.2, 9mm.3), taking CO luminosities and line ratios from their Table 1, and drawing ancillary information from Table 1 of \citet{boogaard+20}.

\citet{lenkic+23} derive $r_{31}$ for a sample of eight $z\sim0.1$ galaxies selected to be analogous to high redshift ($z\sim1-2$) main-sequence galaxies. We draw CO luminosities and line ratios from their Table 3 and ancillary information from \citet[][Table 1]{fisher+19}.

\subsection{Starbursts}

\citet{papadopoulos+12} report line ratios for a large sample of IR selected galaxies (primarily ULIRGs) in the nearby universe. We retrieve $r_{31}$ values for 68 sources and $r_{32}$ values for 31 sources based on their data. We take CO(1--0) luminosities, $r_{21}$ and $r_{31}$ values, and IR luminosities from their Table 7. As $r_{32}$ is not explicitly reported we use $r_{32} = r_{31}/r_{21}$. 

\citet{montoya-arroyave+23} report line ratios for an additional sample of ULIRGs. We take SFRs from their Table 1 and CO luminosities from their Table D.1. We derive line ratios from the reported CO luminosities. Because the \citet{montoya-arroyave+23} CO(1--0) luminosities are derived from interferometric observations, we include $r_{31}$ measurements only for galaxies at $z>0.03$ and where the CO(1--0) maximum-recoverable scale is larger than 15'', in order to ensure the full CO(1--0) flux is recovered. In total we use 11 $r_{31}$ and 29 $r_{32}$ values from their sample.

We cross matched these (U)LIRG samples with the Great Observatories All-sky LIRG Survey \citep[GOALS][]{diaz-santos+17,diaz-santos+13} in order to obtain half-light radii based on the GOALS 70 micron photometry.

\citet{sharon+16}, \citet{friascastillo+23}, and \citet{taylor+25} provide $r_{31}$ values for a total of 32 of high-$z$ SMGs based on a combination of new and literature observations. We draw CO luminosities, line ratios, and ancillary information from Table 2 of Sharon et al., Tables 2 and 4 of Frias Castillo et al., and Tables 1 and A1 of Taylor et al. preferring data from the most recent source when a galaxy appears multiple times.

\section{Further Cloud Model Comparisons}\label{ap:morefits}

\begin{figure*}
    \centering
    \includegraphics[width=\textwidth]{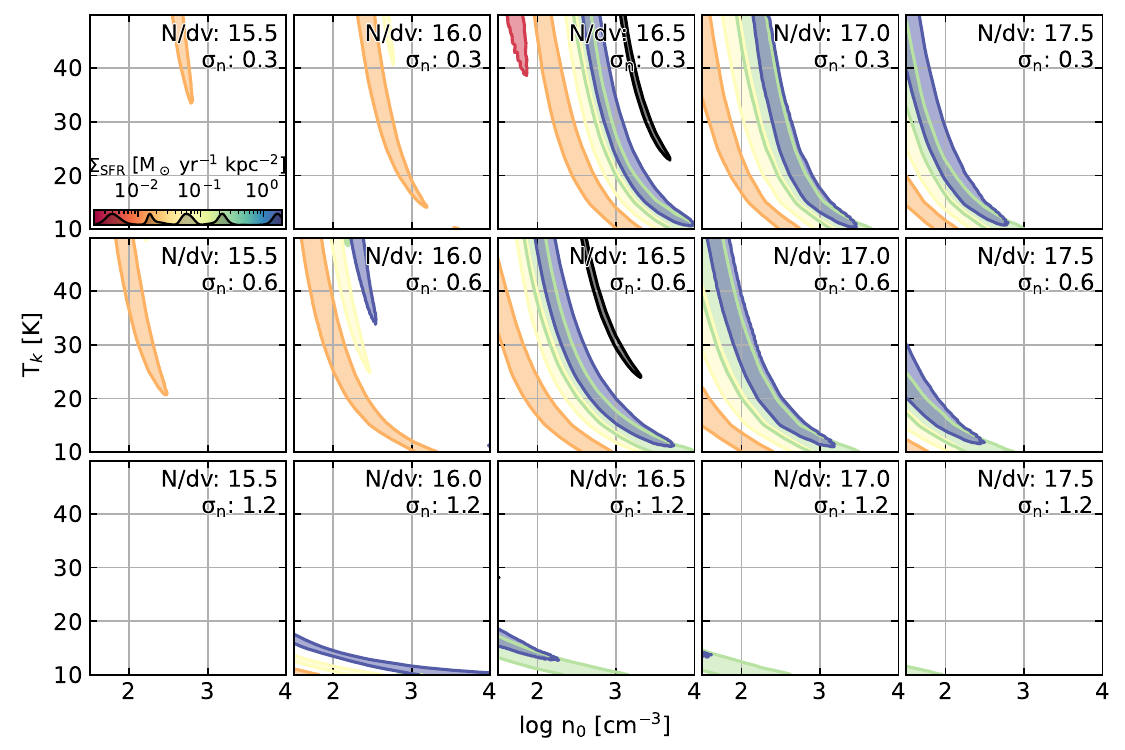}
    \caption{Filled contours represent the range of $n_0$ and $T_k$ for models producing line ratios within $1\sigma$ of the median observed line ratios in bins of $\Sigma_{\rm SFR}$ (colored contours) and for ULIRGs (black countours). The bins are color coded according to their $\Sigma_{\rm SFR}$. Each panel shows results for a fixed model $N/dv$ and $\sigma$. The middle panel in the upper row corresponds to the inset of Figure~\ref{fig:rrmodels}.}
    \label{fig:morefits}
\end{figure*}

In Section~\ref{sec:ism_conditions} we limit our discussion primarily to cloud models with $\sigma_n=0.3$~dex and $N/dv=10^{16.5}\,{\rm cm^{-2}\,(km\,s^{-1})^{-1}}$. In Figure~\ref{fig:morefits} we show the $T_k$ and $n_0$ values compatible with our data over a grid in $\sigma_n$ and $N/dv$. Most of these other parameter choices cannot reproduce the full range of observed line ratios found in our sample. Varying $\sigma_n$ between 0.3~dex and 0.6~dex provides another plausible path from the line ratios in the $\Sigma_{\rm SFR}\sim2\,{\rm M_\odot\,yr^{-1}\,kpc^{-2}}$ bin to those in the ULIRG bin.

\bibliography{refs}

\begin{thebibliography}{}
\expandafter\ifx\csname natexlab\endcsname\relax\def\natexlab#1{#1}\fi
\providecommand{\url}[1]{\href{#1}{#1}}

\bibitem[{{Anirudh} {et~al.}(2025){Anirudh}, {Kaasinen}, {Popping}, {Narayanan}, {Garcia}, \& {Valentin-Martinez}}]{anirudh+25}
{Anirudh}, R., {Kaasinen}, M., {Popping}, G., {et~al.} 2025, arXiv e-prints, arXiv:2506.13899

\bibitem[{{Bauermeister} {et~al.}(2013){Bauermeister}, {Blitz}, {Bolatto}, {Bureau}, {Teuben}, {Wong}, \& {Wright}}]{bauermeister+13a}
{Bauermeister}, A., {Blitz}, L., {Bolatto}, A., {et~al.} 2013, \apj, 763, 64

\bibitem[{{Bolatto} {et~al.}(2013){Bolatto}, {Wolfire}, \& {Leroy}}]{bolatto+13}
{Bolatto}, A.~D., {Wolfire}, M., \& {Leroy}, A.~K. 2013, \araa, 51, 207

\bibitem[{{Bolatto} {et~al.}(2015){Bolatto}, {Warren}, {Leroy}, {Tacconi}, {Bouch{\'e}}, {F{\"o}rster Schreiber}, {Genzel}, {Cooper}, {Fisher}, {Combes}, {Garc{\'\i}a-Burillo}, {Burkert}, {Bournaud}, {Weiss}, {Saintonge}, {Wuyts}, \& {Sternberg}}]{bolatto+15}
{Bolatto}, A.~D., {Warren}, S.~R., {Leroy}, A.~K., {et~al.} 2015, \apj, 809, 175

\bibitem[{{Boogaard} {et~al.}(2020){Boogaard}, {van der Werf}, {Weiss}, {Popping}, {Decarli}, {Walter}, {Aravena}, {Bouwens}, {Riechers}, {Gonz{\'a}lez-L{\'o}pez}, {Smail}, {Carilli}, {Kaasinen}, {Daddi}, {Cox}, {D{\'\i}az-Santos}, {Inami}, {Cortes}, \& {Wagg}}]{boogaard+20}
{Boogaard}, L.~A., {van der Werf}, P., {Weiss}, A., {et~al.} 2020, \apj, 902, 109

\bibitem[{{Bournaud} {et~al.}(2015){Bournaud}, {Daddi}, {Wei{\ss}}, {Renaud}, {Mastropietro}, \& {Teyssier}}]{bournaud+15}
{Bournaud}, F., {Daddi}, E., {Wei{\ss}}, A., {et~al.} 2015, \aap, 575, A56

\bibitem[{{Chabrier}(2003)}]{chabrier03}
{Chabrier}, G. 2003, \apjl, 586, L133

\bibitem[{{Daddi} {et~al.}(2010){Daddi}, {Bournaud}, {Walter}, {Dannerbauer}, {Carilli}, {Dickinson}, {Elbaz}, {Morrison}, {Riechers}, {Onodera}, {Salmi}, {Krips}, \& {Stern}}]{daddi+10}
{Daddi}, E., {Bournaud}, F., {Walter}, F., {et~al.} 2010, \apj, 713, 686

\bibitem[{{Daddi} {et~al.}(2015){Daddi}, {Dannerbauer}, {Liu}, {Aravena}, {Bournaud}, {Walter}, {Riechers}, {Magdis}, {Sargent}, {B{\'e}thermin}, {Carilli}, {Cibinel}, {Dickinson}, {Elbaz}, {Gao}, {Gobat}, {Hodge}, \& {Krips}}]{daddi+15}
{Daddi}, E., {Dannerbauer}, H., {Liu}, D., {et~al.} 2015, \aap, 577, A46

\bibitem[{{den Brok} {et~al.}(2025){den Brok}, {Oakes}, {Leroy}, {Koch}, {Usero}, {Rosolowsky}, {Bigiel}, {Sun}, {He}, {Barnes}, {Cao}, {Liang}, {Pan}, {Saito}, {Sarbadhicary}, \& {Williams}}]{denbrok+25}
{den Brok}, J., {Oakes}, E.~K., {Leroy}, A.~K., {et~al.} 2025, arXiv e-prints, arXiv:2506.09125

\bibitem[{{den Brok} {et~al.}(2021){den Brok}, {Chatzigiannakis}, {Bigiel}, {Puschnig}, {Barnes}, {Leroy}, {Jim{\'e}nez-Donaire}, {Usero}, {Schinnerer}, {Rosolowsky}, {Faesi}, {Grasha}, {Hughes}, {Kruijssen}, {Liu}, {Neumann}, {Pety}, {Querejeta}, {Saito}, {Schruba}, \& {Stuber}}]{denbrok+21}
{den Brok}, J.~S., {Chatzigiannakis}, D., {Bigiel}, F., {et~al.} 2021, \mnras, 504, 3221

\bibitem[{{den Brok} {et~al.}(2023){den Brok}, {Leroy}, {Usero}, {Schinnerer}, {Rosolowsky}, {Koch}, {Querejeta}, {Liu}, {Bigiel}, {Barnes}, {Chevance}, {Colombo}, {Dale}, {Glover}, {Jimenez-Donaire}, {Teng}, \& {Williams}}]{denbrok+23b}
{den Brok}, J.~S., {Leroy}, A.~K., {Usero}, A., {et~al.} 2023, \mnras, 526, 6347

\bibitem[{{Deng} {et~al.}(2025){Deng}, {Li}, {Li}, {Liu}, {Ren}, {Athikkat-Eknath}, {de Grijs}, {Eales}, {Eden}, {Iono}, {Jiao}, {Lee}, {Li}, {Saintonge}, {Smith}, {Tang}, {Tsai}, {van der Giessen}, {Williams}, \& {Wu}}]{deng+25}
{Deng}, Y., {Li}, Z., {Li}, Z., {et~al.} 2025, \mnras, 538, 2445

\bibitem[{{D{\'\i}az-Santos} {et~al.}(2013){D{\'\i}az-Santos}, {Armus}, {Charmandaris}, {Stierwalt}, {Murphy}, {Haan}, {Inami}, {Malhotra}, {Meijerink}, {Stacey}, {Petric}, {Evans}, {Veilleux}, {van der Werf}, {Lord}, {Lu}, {Howell}, {Appleton}, {Mazzarella}, {Surace}, {Xu}, {Schulz}, {Sanders}, {Bridge}, {Chan}, {Frayer}, {Iwasawa}, {Melbourne}, \& {Sturm}}]{diaz-santos+13}
{D{\'\i}az-Santos}, T., {Armus}, L., {Charmandaris}, V., {et~al.} 2013, \apj, 774, 68

\bibitem[{{D{\'\i}az-Santos} {et~al.}(2017){D{\'\i}az-Santos}, {Armus}, {Charmandaris}, {Lu}, {Stierwalt}, {Stacey}, {Malhotra}, {van der Werf}, {Howell}, {Privon}, {Mazzarella}, {Goldsmith}, {Murphy}, {Barcos-Mu{\~n}oz}, {Linden}, {Inami}, {Larson}, {Evans}, {Appleton}, {Iwasawa}, {Lord}, {Sanders}, \& {Surace}}]{diaz-santos+17}
---. 2017, \apj, 846, 32

\bibitem[{{Downes} \& {Solomon}(1998)}]{downes+98}
{Downes}, D., \& {Solomon}, P.~M. 1998, \apj, 507, 615

\bibitem[{{Dunne} {et~al.}(2022){Dunne}, {Maddox}, {Papadopoulos}, {Ivison}, \& {Gomez}}]{dunne+22}
{Dunne}, L., {Maddox}, S.~J., {Papadopoulos}, P.~P., {Ivison}, R.~J., \& {Gomez}, H.~L. 2022, \mnras, 517, 962

\bibitem[{{Egusa} {et~al.}(2022){Egusa}, {Gao}, {Morokuma-Matsui}, {Liu}, \& {Maeda}}]{egusa+22}
{Egusa}, F., {Gao}, Y., {Morokuma-Matsui}, K., {Liu}, G., \& {Maeda}, F. 2022, \apj, 935, 64

\bibitem[{{Fisher} {et~al.}(2019){Fisher}, {Bolatto}, {White}, {Glazebrook}, {Abraham}, \& {Obreschkow}}]{fisher+19}
{Fisher}, D.~B., {Bolatto}, A.~D., {White}, H., {et~al.} 2019, \apj, 870, 46

\bibitem[{{Frias Castillo} {et~al.}(2023){Frias Castillo}, {Hodge}, {Rybak}, {van der Werf}, {Smail}, {Birkin}, {Chen}, {Chapman}, {Hill}, {Lagos}, {Liao}, {da Cunha}, {Calistro Rivera}, {Chen}, {Jim{\'e}nez-Andrade}, {Murphy}, {Scott}, {Swinbank}, {Walter}, {Ivison}, \& {Dannerbauer}}]{friascastillo+23}
{Frias Castillo}, M., {Hodge}, J., {Rybak}, M., {et~al.} 2023, \apj, 945, 128

\bibitem[{{Gao} \& {Solomon}(2004)}]{gao+04}
{Gao}, Y., \& {Solomon}, P.~M. 2004, \apj, 606, 271

\bibitem[{{Gong} {et~al.}(2020){Gong}, {Ostriker}, {Kim}, \& {Kim}}]{gong+20}
{Gong}, M., {Ostriker}, E.~C., {Kim}, C.-G., \& {Kim}, J.-G. 2020, \apj, 903, 142

\bibitem[{{Greve} {et~al.}(2014){Greve}, {Leonidaki}, {Xilouris}, {Wei{\ss}}, {Zhang}, {van der Werf}, {Aalto}, {Armus}, {D{\'{\i}}az-Santos}, {Evans}, {Fischer}, {Gao}, {Gonz{\'a}lez-Alfonso}, {Harris}, {Henkel}, {Meijerink}, {Naylor}, {Smith}, {Spaans}, {Stacey}, {Veilleux}, \& {Walter}}]{greve+14}
{Greve}, T.~R., {Leonidaki}, I., {Xilouris}, E.~M., {et~al.} 2014, \apj, 794, 142

\bibitem[{{Harrington} {et~al.}(2021){Harrington}, {Weiss}, {Yun}, {Magnelli}, {Sharon}, {Leung}, {Vishwas}, {Wang}, {Frayer}, {Jim{\'e}nez-Andrade}, {Liu}, {Garc{\'\i}a}, {Romano-D{\'\i}az}, {Frye}, {Jarugula}, {B{\u{a}}descu}, {Berman}, {Dannerbauer}, {D{\'\i}az-S{\'a}nchez}, {Grassitelli}, {Kamieneski}, {Kim}, {Kirkpatrick}, {Lowenthal}, {Messias}, {Puschnig}, {Stacey}, {Torne}, \& {Bertoldi}}]{harrington+21}
{Harrington}, K.~C., {Weiss}, A., {Yun}, M.~S., {et~al.} 2021, \apj, 908, 95

\bibitem[{{Hogg} {et~al.}(2010){Hogg}, {Bovy}, \& {Lang}}]{hogg+10}
{Hogg}, D.~W., {Bovy}, J., \& {Lang}, D. 2010, arXiv e-prints, arXiv:1008.4686

\bibitem[{{Janowiecki} {et~al.}(2017){Janowiecki}, {Catinella}, {Cortese}, {Saintonge}, {Brown}, \& {Wang}}]{janowiecki+17}
{Janowiecki}, S., {Catinella}, B., {Cortese}, L., {et~al.} 2017, \mnras, 466, 4795

\bibitem[{{Kamenetzky} {et~al.}(2014){Kamenetzky}, {Rangwala}, {Glenn}, {Maloney}, \& {Conley}}]{kamenetzky+14}
{Kamenetzky}, J., {Rangwala}, N., {Glenn}, J., {Maloney}, P.~R., \& {Conley}, A. 2014, \apj, 795, 174

\bibitem[{{Kamenetzky} {et~al.}(2016){Kamenetzky}, {Rangwala}, {Glenn}, {Maloney}, \& {Conley}}]{kamenetzky+16}
---. 2016, \apj, 829, 93

\bibitem[{{Keenan} {et~al.}(2025){Keenan}, {Marrone}, \& {Keating}}]{keenan+24b}
{Keenan}, R.~P., {Marrone}, D.~P., \& {Keating}, G.~K. 2025, \apj, 979, 228

\bibitem[{{Keenan} {et~al.}(2024){Keenan}, {Marrone}, {Keating}, {Mayer}, {Bays}, {Downey}, {Dunn}, {Flores}, {Folkers}, {Forbes}, {Guvenen}, {Holmstedt}, {Moulton}, \& {Sullivan}}]{keenan+24a}
{Keenan}, R.~P., {Marrone}, D.~P., {Keating}, G.~K., {et~al.} 2024, \apj, 975, 150

\bibitem[{{Kennicutt} \& {De Los Reyes}(2021)}]{kennicutt+21}
{Kennicutt}, Robert~C., J., \& {De Los Reyes}, M. A.~C. 2021, \apj, 908, 61

\bibitem[{{Kennicutt} \& {Evans}(2012)}]{kennicutt+12}
{Kennicutt}, R.~C., \& {Evans}, N.~J. 2012, \araa, 50, 531

\bibitem[{{Koda} {et~al.}(2025){Koda}, {Egusa}, {Hirota}, {Lee}, {Sawada}, \& {Maeda}}]{koda+25}
{Koda}, J., {Egusa}, F., {Hirota}, A., {et~al.} 2025, arXiv e-prints, arXiv:2505.08876

\bibitem[{{Komugi} {et~al.}(2007){Komugi}, {Kohno}, {Tosaki}, {Nakanishi}, {Onodera}, {Egusa}, \& {Sofue}}]{komugi+07}
{Komugi}, S., {Kohno}, K., {Tosaki}, T., {et~al.} 2007, \pasj, 59, 55

\bibitem[{{Komugi} {et~al.}(2025){Komugi}, {Sawada}, {Koda}, {Egusa}, {Maeda}, {Hirota}, \& {Lee}}]{komugi+25}
{Komugi}, S., {Sawada}, T., {Koda}, J., {et~al.} 2025, \apj, 980, 126

\bibitem[{{Lamperti} {et~al.}(2020){Lamperti}, {Saintonge}, {Koss}, {Viti}, {Wilson}, {He}, {Shimizu}, {Greve}, {Mushotzky}, {Treister}, {Kramer}, {Sanders}, {Schawinski}, \& {Tacconi}}]{lamperti+20}
{Lamperti}, I., {Saintonge}, A., {Koss}, M., {et~al.} 2020, \apj, 889, 103

\bibitem[{{Lenki{\'c}} {et~al.}(2023){Lenki{\'c}}, {Bolatto}, {Fisher}, {Abraham}, {Glazebrook}, {Herrera-Camus}, {Levy}, {Obreschkow}, \& {Volpert}}]{lenkic+23}
{Lenki{\'c}}, L., {Bolatto}, A.~D., {Fisher}, D.~B., {et~al.} 2023, \apj, 945, 9

\bibitem[{{Leroy} {et~al.}(2017){Leroy}, {Usero}, {Schruba}, {Bigiel}, {Kruijssen}, {Kepley}, {Blanc}, {Bolatto}, {Cormier}, {Gallagher}, {Hughes}, {Jim{\'e}nez-Donaire}, {Rosolowsky}, \& {Schinnerer}}]{leroy+17}
{Leroy}, A.~K., {Usero}, A., {Schruba}, A., {et~al.} 2017, \apj, 835, 217

\bibitem[{{Leroy} {et~al.}(2021){Leroy}, {Schinnerer}, {Hughes}, {Rosolowsky}, {Pety}, {Schruba}, {Usero}, {Blanc}, {Chevance}, {Emsellem}, {Faesi}, {Herrera}, {Liu}, {Meidt}, {Querejeta}, {Saito}, {Sandstrom}, {Sun}, {Williams}, {Anand}, {Barnes}, {Behrens}, {Belfiore}, {Benincasa}, {Be{\v{s}}li{\'c}}, {Bigiel}, {Bolatto}, {den Brok}, {Cao}, {Chandar}, {Chastenet}, {Chiang}, {Congiu}, {Dale}, {Deger}, {Eibensteiner}, {Egorov}, {Garc{\'\i}a-Rodr{\'\i}guez}, {Glover}, {Grasha}, {Henshaw}, {Ho}, {Kepley}, {Kim}, {Klessen}, {Kreckel}, {Koch}, {Kruijssen}, {Larson}, {Lee}, {Lopez}, {Machado}, {Mayker}, {McElroy}, {Murphy}, {Ostriker}, {Pan}, {Pessa}, {Puschnig}, {Razza}, {S{\'a}nchez-Bl{\'a}zquez}, {Santoro}, {Sardone}, {Scheuermann}, {Sliwa}, {Sormani}, {Stuber}, {Thilker}, {Turner}, {Utomo}, {Watkins}, \& {Whitmore}}]{leroy+21}
{Leroy}, A.~K., {Schinnerer}, E., {Hughes}, A., {et~al.} 2021, \apjs, 257, 43

\bibitem[{{Leroy} {et~al.}(2022){Leroy}, {Rosolowsky}, {Usero}, {Sandstrom}, {Schinnerer}, {Schruba}, {Bolatto}, {Sun}, {Barnes}, {Belfiore}, {Bigiel}, {den Brok}, {Cao}, {Chiang}, {Chevance}, {Dale}, {Eibensteiner}, {Faesi}, {Glover}, {Hughes}, {Jim{\'e}nez Donaire}, {Klessen}, {Koch}, {Kruijssen}, {Liu}, {Meidt}, {Pan}, {Pety}, {Puschnig}, {Querejeta}, {Saito}, {Sardone}, {Watkins}, {Weiss}, \& {Williams}}]{leroy+22}
{Leroy}, A.~K., {Rosolowsky}, E., {Usero}, A., {et~al.} 2022, \apj, 927, 149

\bibitem[{{Li} {et~al.}(2020){Li}, {Li}, {Smith}, {Wilson}, {Gao}, {Eales}, {Ao}, {Bureau}, {Chung}, {Davis}, {de Grijs}, {Eden}, {He}, {Hughes}, {Jiang}, {Kemper}, {Lamperti}, {Lee}, {Lee}, {Micha{\l}owski}, {Parsons}, {Ragan}, {Scicluna}, {Shi}, {Tang}, {Tomi{\v{c}}i{\'c}}, {Viaene}, {Williams}, \& {Zhu}}]{li+20}
{Li}, Z., {Li}, Z., {Smith}, M. W.~L., {et~al.} 2020, \mnras, 492, 195

\bibitem[{{Liu} {et~al.}(2015){Liu}, {Gao}, {Isaak}, {Daddi}, {Yang}, {Lu}, \& {van der Werf}}]{liu+15}
{Liu}, D., {Gao}, Y., {Isaak}, K., {et~al.} 2015, \apjl, 810, L14

\bibitem[{{Liu} {et~al.}(2021){Liu}, {Daddi}, {Schinnerer}, {Saito}, {Leroy}, {Silverman}, {Valentino}, {Magdis}, {Gao}, {Jin}, {Puglisi}, \& {Groves}}]{liu+21}
{Liu}, D., {Daddi}, E., {Schinnerer}, E., {et~al.} 2021, \apj, 909, 56

\bibitem[{{Maeda} {et~al.}(2022){Maeda}, {Egusa}, {Ohta}, {Fujimoto}, {Habe}, \& {Asada}}]{maeda+23}
{Maeda}, F., {Egusa}, F., {Ohta}, K., {et~al.} 2022, \apj, 926, 96

\bibitem[{{Mao} {et~al.}(2010){Mao}, {Schulz}, {Henkel}, {Mauersberger}, {Muders}, \& {Dinh-V-Trung}}]{mao+10}
{Mao}, R.-Q., {Schulz}, A., {Henkel}, C., {et~al.} 2010, \apj, 724, 1336

\bibitem[{{Miura} {et~al.}(2014){Miura}, {Kohno}, {Tosaki}, {Espada}, {Hirota}, {Komugi}, {Okumura}, {Kuno}, {Muraoka}, {Onodera}, {Nakanishi}, {Sawada}, {Kaneko}, {Minamidani}, {Fujii}, \& {Kawabe}}]{miura+14}
{Miura}, R.~E., {Kohno}, K., {Tosaki}, T., {et~al.} 2014, \apj, 788, 167

\bibitem[{{Montoya Arroyave} {et~al.}(2023){Montoya Arroyave}, {Cicone}, {Makroleivaditi}, {Weiss}, {Lundgren}, {Severgnini}, {De Breuck}, {Baumschlager}, {Schimek}, {Shen}, \& {Aravena}}]{montoya-arroyave+23}
{Montoya Arroyave}, I., {Cicone}, C., {Makroleivaditi}, E., {et~al.} 2023, \aap, 673, A13

\bibitem[{{Morokuma-Matsui} \& {Muraoka}(2017)}]{moromkuma-matsui+17}
{Morokuma-Matsui}, K., \& {Muraoka}, K. 2017, \apj, 837, 137

\bibitem[{{Narayanan} {et~al.}(2005){Narayanan}, {Groppi}, {Kulesa}, \& {Walker}}]{narayanan+05}
{Narayanan}, D., {Groppi}, C.~E., {Kulesa}, C.~A., \& {Walker}, C.~K. 2005, \apj, 630, 269

\bibitem[{{Narayanan} \& {Krumholz}(2014)}]{narayanan+14}
{Narayanan}, D., \& {Krumholz}, M.~R. 2014, \mnras, 442, 1411

\bibitem[{{Papadopoulos} {et~al.}(2012){Papadopoulos}, {van der Werf}, {Xilouris}, {Isaak}, {Gao}, \& {M{\"u}hle}}]{papadopoulos+12}
{Papadopoulos}, P.~P., {van der Werf}, P.~P., {Xilouris}, E.~M., {et~al.} 2012, \mnras, 426, 2601

\bibitem[{{Pe{\~n}aloza} {et~al.}(2018){Pe{\~n}aloza}, {Clark}, {Glover}, \& {Klessen}}]{penaloza+18}
{Pe{\~n}aloza}, C.~H., {Clark}, P.~C., {Glover}, S. C.~O., \& {Klessen}, R.~S. 2018, \mnras, 475, 1508

\bibitem[{{Pe{\~n}aloza} {et~al.}(2017){Pe{\~n}aloza}, {Clark}, {Glover}, {Shetty}, \& {Klessen}}]{penaloza+17}
{Pe{\~n}aloza}, C.~H., {Clark}, P.~C., {Glover}, S. C.~O., {Shetty}, R., \& {Klessen}, R.~S. 2017, \mnras, 465, 2277

\bibitem[{{Riechers} {et~al.}(2020){Riechers}, {Boogaard}, {Decarli}, {Gonz{\'a}lez-L{\'o}pez}, {Smail}, {Walter}, {Aravena}, {Carilli}, {Cortes}, {Cox}, {D{\'\i}az-Santos}, {Hodge}, {Inami}, {Ivison}, {Kaasinen}, {Wagg}, {Wei{\ss}}, \& {van der Werf}}]{riechers+20}
{Riechers}, D.~A., {Boogaard}, L.~A., {Decarli}, R., {et~al.} 2020, \apjl, 896, L21

\bibitem[{{Saintonge} \& {Catinella}(2022)}]{saintonge+22}
{Saintonge}, A., \& {Catinella}, B. 2022, \araa, 60, 319

\bibitem[{{Saintonge} {et~al.}(2017){Saintonge}, {Catinella}, {Tacconi}, {Kauffmann}, {Genzel}, {Cortese}, {Dav{\'e}}, {Fletcher}, {Graci{\'a}-Carpio}, {Kramer}, {Heckman}, {Janowiecki}, {Lutz}, {Rosario}, {Schiminovich}, {Schuster}, {Wang}, {Wuyts}, {Borthakur}, {Lamperti}, \& {Roberts-Borsani}}]{saintonge+17}
{Saintonge}, A., {Catinella}, B., {Tacconi}, L.~J., {et~al.} 2017, \apjs, 233, 22

\bibitem[{{Salim} {et~al.}(2016){Salim}, {Lee}, {Janowiecki}, {da Cunha}, {Dickinson}, {Boquien}, {Burgarella}, {Salzer}, \& {Charlot}}]{salim+16}
{Salim}, S., {Lee}, J.~C., {Janowiecki}, S., {et~al.} 2016, \apjs, 227, 2

\bibitem[{{Sargent} {et~al.}(2014){Sargent}, {Daddi}, {B{\'e}thermin}, {Aussel}, {Magdis}, {Hwang}, {Juneau}, {Elbaz}, \& {da Cunha}}]{sargent+14}
{Sargent}, M.~T., {Daddi}, E., {B{\'e}thermin}, M., {et~al.} 2014, \apj, 793, 19

\bibitem[{{Sawada} {et~al.}(2001){Sawada}, {Hasegawa}, {Handa}, {Morino}, {Oka}, {Booth}, {Bronfman}, {Hayashi}, {Luna Castellanos}, {Nyman}, {Sakamoto}, {Seta}, {Shaver}, {Sorai}, \& {Usuda}}]{sawada+01}
{Sawada}, T., {Hasegawa}, T., {Handa}, T., {et~al.} 2001, \apjs, 136, 189

\bibitem[{{Schinnerer} \& {Leroy}(2024)}]{schinnerer+24}
{Schinnerer}, E., \& {Leroy}, A.~K. 2024, arXiv e-prints, arXiv:2403.19843

\bibitem[{{Sharon} {et~al.}(2016){Sharon}, {Riechers}, {Hodge}, {Carilli}, {Walter}, {Wei{\ss}}, {Knudsen}, \& {Wagg}}]{sharon+16}
{Sharon}, C.~E., {Riechers}, D.~A., {Hodge}, J., {et~al.} 2016, \apj, 827, 18

\bibitem[{{Speagle} {et~al.}(2014){Speagle}, {Steinhardt}, {Capak}, \& {Silverman}}]{speagle+14}
{Speagle}, J.~S., {Steinhardt}, C.~L., {Capak}, P.~L., \& {Silverman}, J.~D. 2014, \apjs, 214, 15

\bibitem[{{Sun} {et~al.}(2020){Sun}, {Leroy}, {Schinnerer}, {Hughes}, {Rosolowsky}, {Querejeta}, {Schruba}, {Liu}, {Saito}, {Herrera}, {Faesi}, {Usero}, {Pety}, {Kruijssen}, {Ostriker}, {Bigiel}, {Blanc}, {Bolatto}, {Boquien}, {Chevance}, {Dale}, {Deger}, {Emsellem}, {Glover}, {Grasha}, {Groves}, {Henshaw}, {Jimenez-Donaire}, {Kim}, {Klessen}, {Kreckel}, {Lee}, {Meidt}, {Sandstrom}, {Sardone}, {Utomo}, \& {Williams}}]{sun+20}
{Sun}, J., {Leroy}, A.~K., {Schinnerer}, E., {et~al.} 2020, \apjl, 901, L8

\bibitem[{{Taylor} {et~al.}(2025){Taylor}, {Swinbank}, {Smail}, {Puglisi}, {Birkin}, {Dudzevi{\v{c}}i{\={u}}t{\.{e}}}, {Chen}, {Ikarashi}, {Frias Castillo}, {Wei{\ss}}, {Li}, {Chapman}, {Jansen}, {Jim{\'e}nez-Andrade}, {Morabito}, {Murphy}, {Rybak}, \& {van der Werf}}]{taylor+25}
{Taylor}, D.~J., {Swinbank}, A.~M., {Smail}, I., {et~al.} 2025, \mnras, 536, 1149

\bibitem[{{Valentino} {et~al.}(2020){Valentino}, {Daddi}, {Puglisi}, {Magdis}, {Liu}, {Kokorev}, {Cortzen}, {Madden}, {Aravena}, {G{\'o}mez-Guijarro}, {Lee}, {Le Floc'h}, {Gao}, {Gobat}, {Bournaud}, {Dannerbauer}, {Jin}, {Dickinson}, {Kartaltepe}, \& {Sanders}}]{valentino+20}
{Valentino}, F., {Daddi}, E., {Puglisi}, A., {et~al.} 2020, \aap, 641, A155

\bibitem[{{van der Tak} {et~al.}(2007){van der Tak}, {Black}, {Sch{\"o}ier}, {Jansen}, \& {van Dishoeck}}]{vandertak+07}
{van der Tak}, F.~F.~S., {Black}, J.~H., {Sch{\"o}ier}, F.~L., {Jansen}, D.~J., \& {van Dishoeck}, E.~F. 2007, \aap, 468, 627

\bibitem[{{Wei{\ss}} {et~al.}(2005){Wei{\ss}}, {Walter}, \& {Scoville}}]{weiss+05}
{Wei{\ss}}, A., {Walter}, F., \& {Scoville}, N.~Z. 2005, \aap, 438, 533

\bibitem[{{Yajima} {et~al.}(2021){Yajima}, {Sorai}, {Miyamoto}, {Muraoka}, {Kuno}, {Kaneko}, {Takeuchi}, {Yasuda}, {Tanaka}, {Morokuma-Matsui}, \& {Kobayashi}}]{yajima+21}
{Yajima}, Y., {Sorai}, K., {Miyamoto}, Y., {et~al.} 2021, \pasj, 73, 257

\bibitem[{{Yao} {et~al.}(2003){Yao}, {Seaquist}, {Kuno}, \& {Dunne}}]{yao+03}
{Yao}, L., {Seaquist}, E.~R., {Kuno}, N., \& {Dunne}, L. 2003, \apj, 588, 771

\end{thebibliography}

\end{document}